\documentclass[twocolumn]{aastex701}
\usepackage{graphicx} % Required for inserting images
\usepackage{physics}
\usepackage{ragged2e}
\usepackage{rotating} % for \rotatebox
\usepackage{amsmath}
\usepackage[version=4]{mhchem}
\begin{document}
\title{CO in MASsive Spirals (CO-MASS): an IRAM 30m CO emission line survey of the CGM-MASS sample}

\correspondingauthor{Jiang-Tao Li}
\email{pandataotao@gmail.com}

%\author{Yu Huang, Jiang-Tao Li, Yan Jiang, Ping Zhou}

\author[0009-0007-5968-2236]{Yu Huang}
\affiliation{Purple Mountain Observatory, Chinese Academy of Sciences, 10 Yuanhua Road, Nanjing 210023, People’s Republic of China}
\affiliation{School of Astronomy and Space Science, University of Science and Technology of China, Hefei 230026, People's Republic of China}
\email{yuhuang@pmo.ac.cn}

\author[0000-0001-6239-3821]{Jiang-Tao Li}
\affiliation{Purple Mountain Observatory, Chinese Academy of Sciences, 10 Yuanhua Road, Nanjing 210023, People’s Republic of China}
\email{pandataotao@gmail.com}

\author[0009-0003-3907-5077]{Yan Jiang}
\affiliation{School of Physics and Astronomy, China West Normal University, No. 1 Shida Road, Nanchong 637002, People’s Republic of China}
\email{yanjiang@cwnu.edu.cn}

\author[0000-0002-5683-822X]{Ping Zhou}
\affiliation{Key Laboratory of Modern Astronomy and Astrophysics, Nanjing University, Ministry of Education, Nanjing 210093, People's Republic of China}
\affiliation{Key Laboratory of Radio Astronomy, Chinese Academy of Sciences, Nanjing 210023, People's Republic of China}
\email{pingzhou@nju.edu.cn}

\author[0009-0006-3887-8988]{Jianghui Xu}
\affiliation{Department of Astronomy, University of Science and Technology of China, Hefei, Anhui 230026, People's Republic of China}
\affiliation{School of Astronomy and Space Science, University of Science and Technology of China, Hefei 230026, People's Republic of China}
\email{paczynski@mail.ustc.edu.cn}

\author[0000-0002-3286-5346]{Li-Yuan Lu}
\affiliation{Department of Physics and Astronomy, Qinghai University, 251 Ningda Road, Xining, 810016, China}
\affiliation{Research Center of Astronomy, Qinghai University, 251 Ningda Road, Xining, 810016, China}
\email{lylu@qhu.edu.cn}

\author[0000-0001-7254-219X]{Yang Yang}
\affiliation{Xiangtan University, Xiangtan 411105, Hunan, P.R. China}
\email{yangyang.astro@gmail.com}

\begin{abstract}

There exist extremely massive spiral galaxies in isolated environments, with stellar masses several times that of the Milky Way, yet their star formation rates (SFRs) are comparable to or even lower than that of the Milky Way. In this paper, we investigate the molecular gas properties of such galaxies to better understand the origin of their low SFRs. We present IRAM 30m CO observations of five extremely massive spirals from the CGM-MASS sample. We compare their star formation efficiencies (SFEs) with the Kennicutt-Schmidt relation and find that these massive spirals generally exhibit low efficiency in converting molecular gas into stars. We further compare their molecular gas masses with their atomic gas and stellar masses, and also include the CHANG-ES sample galaxies observed with the IRAM 30m telescope in a similar manner for comparison. Our sample galaxies show low efficiency in converting atomic to molecular gas and have lower molecular gas fractions, suggesting that their suppressed star formation stems from both limited gas supply and inefficient star formation. Considering potential cold gas sources in massive spirals, we argue that their current reservoirs likely originate from past starburst or merger events rather than ongoing accretion in present isolated environments. Finally, we examine the location of these galaxies on the baryonic Tully-Fisher relation, finding them baryon-deficient and deviating from the trend of lower-mass galaxies. This suggests either a significant undetected baryonic component or a flattening/turnover of the relation at the high-mass end, consistent with the stellar mass–halo mass relation.

\end{abstract}

\keywords{\uat{Interstellar medium}{847} --- \uat{Interstellar molecules}{849} --- \uat{Spiral galaxies}{1560} --- \uat{Star formation}{1569}}

\section{Introduction}\label{sec:Intro}

Molecular gas plays a crucial role in galaxy evolution. As the primary fuel for star formation, it is essential for understanding how stars form and evolve within galaxies \citetext{e.g., \citealt{bolatto13}}. Observational studies have revealed a strong correlation between the star formation rate (SFR) and the gas content of galaxies. Early studies found that the SFR scales with the total neutral gas content (molecular hydrogen $\rm H_2$ plus atomic hydrogen \ion{H}{1}), indicating that regions with higher gas surface densities tend to host more active star formation \citep{kennicutt98, schmidt59}. This relationship is commonly referred to as the Kennicutt–Schmidt (K–S) law. Subsequent studies refined this picture by showing that the SFR correlates more tightly with molecular gas alone \citetext{e.g., \citealt{Leroy08, bigiel08}}, highlighting the central role of molecular gas in regulating star formation across diverse galactic environments.

Massive spiral galaxies ($\mathrm{log}_{10}(M_\star/\rm M_\odot)>10.5$) may exhibit star formation properties distinct from those of lower-mass spirals well defined with the K-S law \citetext{e.g., \citealt{schreiber15, Ogle2016}}. These differences likely arise from their unique formation histories and relatively quiescent environments. Unlike massive ellipticals, which often result from multiple major mergers, massive spirals must have evolved through more secular processes, as spiral structures are typically disrupted by major mergers \citetext{e.g., \citealt{toomre72, hopkins09}}. Although rare, some exceptionally massive spiral galaxies—reaching stellar masses several times that of the Milky Way (MW)—have been identified in isolated environments (e.g., \citealt{Ogle2016,jiangtao17}). Interestingly, despite their large stellar masses, these systems exhibit specific star formation rates (sSFR = SFR/$M_{\star}$) that are comparable to or even lower than that of the MW \citetext{e.g., \citealt{jiangtao16}}. The existence of such galaxies raises fundamental questions: How did they assemble such massive stellar components? Why is their current star formation so inefficient? And what mechanisms are responsible for regulating or suppressing star formation in these extreme systems?

Previous studies have investigated the molecular gas content in a few ultra-massive spiral galaxies, such as NGC~5908 \citep{jiangtao19} and NGC~4594 \citep{Yan25}. However, these galaxies present complications that limit their suitability as benchmarks for understanding star formation in massive spirals. NGC~5908 is likely influenced by \ion{H}{1} gas transfer from its nearby companion NGC~5905, while NGC~4594 (the Sombrero Galaxy) hosts a powerful radio jet, suggesting possible AGN-driven quenching \citep{yang24}. To better isolate the intrinsic properties of molecular gas and star formation in massive spiral galaxies, a cleaner and more representative sample is needed.

The Circum-Galactic Medium of MASsive Spirals (CGM-MASS) survey identified five isolated, massive spiral galaxies selected based on a set of well-defined criteria \citep{jiangtao16}. These include: maximum gas rotation velocity $v_{\rm rot} \gtrsim 300~\rm km~s^{-1}$, low Galactic foreground absorption column density $N_{\rm H} < 10^{21}~\rm cm^{-2}$, distance $< 100~\rm Mpc$, stellar mass $M_{\star} \gtrsim 2 \times 10^{11}~\mathrm{M_{\odot}}$, virial radius angular size $r_{200} < 35^{\prime}$, and the absence of bright companions within $10^{\prime}$. The angular sizes of $35^{\prime}$ and $10^{\prime}$ correspond to physical sizes of about 700 and 200 kpc, respectively, at a distance of 70 Mpc. These galaxies provide ideal laboratories for studying the molecular gas content in truly isolated (except for NGC~5908, which has a massive companion NGC~5905 slightly larger than $10^{\prime}$ away), massive spiral systems.

The CO in MASsive Spirals (CO-MASS) project was initiated to study the molecular gas content in the CGM-MASS galaxy sample. A primary goal is to investigate the star formation efficiency (SFE) in these systems and determine whether their relatively weak star formation activity is due to a deficiency of molecular gas or to the presence of quenching mechanisms that suppress star formation within molecular clouds \citetext{e.g., \citealt{kennicutt98, bigiel08, Leroy08, martig09}}. 

%Another key motivation stems from the observed discrepancy between the total baryonic mass—computed as the sum of stellar mass and hot gas mass—and the value predicted by the baryonic Tully-Fisher relation (BTFR) \citep{jiangtao17}. By incorporating the cold gas component into the baryonic mass budget, we aim to evaluate whether this apparent deviation reflects a turnover or flattening in the BTFR at the high-mass end, or whether it indicates the existence of previously unaccounted baryonic components.

In this paper, we first analyze the IRAM 30m CO emission line data of the CGM-MASS galaxies in \S\ref{sec:Observation}. We then present the derived molecular gas properties in \S\ref{sec:result}, followed by a discussion of several key scientific topics in \S\ref{sec:discussion}, including the SFE, the relative content and origin of the molecular gas, and the baryonic Tully-Fisher relation (BTFR). Our main results and conclusions are summarized in \S\ref{sec:conclusion}. The relevant physical parameters of the sample galaxies are compiled from previous studies \citep{jiangtao17, jiangtao19, springob05} and are listed in Table~\ref{table:parameters}. Unless otherwise stated, all quoted uncertainties represent the $1\sigma$ confidence level. 

\renewcommand{\arraystretch}{1.2}
\setlength{\tabcolsep}{7.5pt}
\begin{table*}
\centering
\caption{ Parameters of the CGM-MASS Sample Galaxies } 
\begin{tabular}{cccccccccc}
\hline\hline
Name & Dist  & Size & Type & $V_{\mathrm{rot}}$ & $M_{\mathrm{HI}}$ & $M_{\star}$ & $M_{\mathrm{hot}}$ &$\dot{M}_{\mathrm{hot}}$ & SFR  \\
 & Mpc & $a~\times~b$ & & $\rm km~s^{-1}$ & $10^{10}\mathrm{M_{\odot}}$ & $10^{11} M_{\odot}$ & $10^{11}\mathrm{M_{\odot}}$  & $M_{\odot}~\mathrm{yr^{-1}}$ & $M_{\odot}~\mathrm{yr^{-1}}$  \\
 \hline
UGCA~145 & 69.3 & $3.1^\prime~\times~0.5^\prime$ & SAbc & 329.1 & 8.14 & $1.47^{+0.01}_{-0.08}$& $1.46^{+0.45}_{-0.29}$  & $0.006^{+0.003}_{-0.002}$ & $2.75~\pm~0.11$  \\
NGC~550 & 93.1  &$1.8^\prime~\times~0.7^\prime$ & SB(s)a & 317.9 & 0.47 & $2.58^{+0.04}_{-0.28}$& $1.98^{+0.82}_{-0.48}$ & $0.007^{+0.004}_{-0.003}$ & $0.38~\pm~0.09$ \\
UGC~12591 & 94.4 &$1.5^\prime~\times~0.6^\prime$ & S0/a & 483.5   & 0.51 & $5.92^{+0.14}_{-0.74}$& $3.08^{+1.41}_{-0.74}$ & $0.062^{+0.041}_{-0.022}$ & $1.17~\pm~0.13$ \\
NGC~669 & 77.8  &$2.3^\prime~\times~0.6^\prime$ & Sab & 356.1   & 0.47 & $3.32^{+0.02}_{-0.17}$& $1.05^{+0.34}_{-0.21}$ & $0.054^{+0.026}_{-0.017}$ & $0.77~\pm~0.07$ \\
NGC~5908 & 51.9  &$3.4^\prime~\times~1.6^\prime$& SA(s)b & 347.5  & 2.8 & $2.56^{+0.02}_{-0.15}$& $0.14^{+0.33}_{-0.06}$ & $0.37(<1.55)$ & $3.81~\pm~0.09$ \\
\hline
\end{tabular}
\\
\label{table:parameters}
\justifying\noindent
\textbf{Notes.} Galaxy parameters. Diameter of major (a) and minor (b) axes are obtained from HyperLeda (\url{http://leda.univ-lyon1.fr/}). The morphology type is obtained from NED(\url{https://ned.ipac.caltech.edu}). $M_{\mathrm{HI}}$ is the atomic gas mass obtained from the \ion{H}{1} 21 cm emission line \citep{springob05}. $V_{\mathrm{rot}}$ is the inclination corrected rotation velocity obtained from \citet{jiangtao16, jiangtao17}. 
$M_{\star}$ is the stellar mass estimated from the K-band magnitude. $M_{\mathrm{hot}}$ and $\dot{M}_{\mathrm{hot}}$ is the total hot gas mass and the integrated radiative cooling rate. SFR is the star formation rate estimated from the \textit{WISE} $22~\rm \mu m$ data. The last four are all obtained from \citet{jiangtao17}.

\end{table*}

\section{Observations and Data Reduction}\label{sec:Observation}

We conducted IRAM 30m CO line observations of five galaxies from the CGM-MASS project, taken between July 2016 and February 2017, primarily under projects 062-16 and 162-16 (PI: Jiang-Tao Li). Additional data for NGC~5908 were obtained later during the 2018B semester through project 063-18 (PI: Jiang-Tao Li). As the NGC~5908 data have already been processed and presented in \citet{jiangtao19}, this paper focuses on the data reduction and analysis of the remaining four galaxies. Our CO line observations were performed using the Eight MIxer Receiver (EMIR) in the E90/E230 dual-band configuration to simultaneously target the $^{12}\mathrm{CO}~J=1\text{--}0$ and $^{12}\mathrm{CO}~J=2\text{--}1$ transitions in Wobbler Switching (WSW) mode \citep{carter12}. The Fast Fourier Transform Spectrometer (FTS) was used as the backend, providing a total bandwidth of 32~GHz and a spectral resolution of 200~kHz. The antenna half-power beam width (HPBW) is $21.4^{\prime\prime}$ at 115~GHz and $10.7^{\prime\prime}$ at 230~GHz, with corresponding beam efficiencies of 78\% and 59\%, respectively. For each galaxy, we selected 6 to 10 positions along the disk for pointed observations. The spatial coverage for UGCA~145 is shown as an example in the left panel of Figure~\ref{fig:I_distance}. Similar figures of the remaining galaxies are presented in the appendix (\S\ref{App1}). 

We used the CLASS package (Continuum and Line Analysis Single-dish Software) of the \href{https://www.iram.fr/IRAMFR/GILDAS/}{GILDAS} suite\footnote{\url{https://www.iram.fr/IRAMFR/GILDAS/}} to reduce the data. The $^{12}\mathrm{CO}~J=1\text{--}0$ and $^{12}\mathrm{CO}~J=2\text{--}1$ spectra were binned to a velocity resolution of $10~\mathrm{km~s^{-1}}$. Baseline subtraction was performed for each spectrum using polynomial fitting, with the polynomial order typically limited to three to avoid overfitting. For positions with detectable emission lines, we fit the spectra using a single- or double-Gaussian model plus a linear continuum within a velocity range of $|v| < 1000~\mathrm{km~s^{-1}}$ relative to the galaxy's systemic velocity, and calculated the integrated line intensities. For non-detections, $3\sigma$ upper limits were given, where the RMS noise was measured over emission-free channels within the same velocity window. The radial profiles of the integrated intensities of the $^{12}\mathrm{CO}~J=1\text{--}0$ and $^{12}\mathrm{CO}~J=2\text{--}1$ transitions were constructed along the major axis of each galaxy. Figure~\ref{fig:I_distance} shows UGCA~145 as an example, with the remaining profiles provided in \S\ref{App1}. Additional spectral fitting results are provided in \S\ref{App2} (Figures~\ref{fig:spectra145}--\ref{fig:spectra12591}).

To compute the line intensity ratio between $^{12}\mathrm{CO}~J=2\text{--}1$ and $^{12}\mathrm{CO}~J=1\text{--}0$, a beam dilution correction was applied to account for the different HPBWs at the two observing frequencies. The correction factor is given by
\[
f_{\mathrm{corr}} = \frac{\theta_{10}^2}{\theta_{21}^2},
\]
where $\theta_{10}$ and $\theta_{21}$ are the HPBWs at 115~GHz and 230~GHz, respectively. The detailed results are summarized in Table~\ref{table:COtable}.

\setlength{\tabcolsep}{10pt}
\begin{table*}
\centering
\caption{Observed and Derived Parameters of the CO Lines} 
\begin{tabular}{cccccccc}
\hline\hline
Region & $d$ & $ I_{12\mathrm{CO_{10}}}$ & $v_{12\mathrm{CO_{10}}}$ & $I_{12\mathrm{CO_{21}}}$ & $v_{12\mathrm{CO_{21}}}$ & $\rm \frac{^{12}CO_{21}}{^{12}CO_{10}} $\\
 & kpc & $\rm K~km~s^{-1}$ & $\rm km~s^{-1}$ & $\rm K~km~s^{-1}$ & $\rm km~s^{-1}$ & \\
 \hline
UGCA~145-1 & $-2.15$ & $8.28~\pm~0.54 $ & $87.4~\pm~4.0$   & $5.91~\pm~0.66$ & $63.9~\pm~5.5$    & $0.17~\pm~0.02$\\
UGCA~145-2 & 2.15  & $11.17~\pm~0.76$ & $-46.78~\pm~6.1$ & $12.07~\pm~0.93$& $-111.92~\pm~6.8$ & $0.27~\pm~0.03$\\
UGCA~145-3 & $-6.38$ & $9.60~\pm~0.51 $ & $170.95~\pm~4.5$ & $8.35~\pm~1.20$ & $117.07~\pm~4.7$  & $0.21~\pm~0.03$\\
UGCA~145-4 & 6.38  & $7.00~\pm~0.39 $ & $-200.12~\pm~2.7$& $2.57~\pm~0.53$ & $-283.12~\pm~2.3$ & $0.09~\pm~0.02$\\
UGCA~145-5 & $-10.81$& $8.90~\pm~0.41 $ & $241.63~\pm~2.8$ & $7.80~\pm~0.84$ & $208.59~\pm~8.0$  & $0.21~\pm~0.02$\\
UGCA~145-6 & 10.81 & $7.19~\pm~0.45 $ & $-240.85~\pm~3.7$& $4.54~\pm~0.97$ & $-347.81\pm~9.2$  & $0.16~\pm~0.04$\\ 
UGCA~145-7 & $-15.04 $& $7.09~\pm~0.43 $ & $287.77~\pm~2.7$ & $2.47~\pm~0.68$ & $263.33\pm~10.4$  & $0.08~\pm~0.02$\\
UGCA~145-8 & 15.04 & $2.52~\pm~0.37 $ & $-295.94~\pm~2.2$& $<~2.23$ & - & $<~0.22$\\ 
UGCA~145-9 & $-22.25$ & $1.94~\pm~0.22 $ & $302.01~\pm~3.2$ & $<~1.12$ & - & $<~0.14$\\ 
UGCA~145-10& 22.52 & $1.33~\pm~0.30 $ & $-280.32~\pm~2.9$& $<~1.55$ & - & $<~0.29$\\
\hline

NGC~550-0 & $-3.17$ & $<~0.78$ & - & $<~1.55$ & - & -\\
NGC~550-1 & $-3.17$ & $<~0.76$ & - & $<~1.46$ & - & -\\
NGC~550-2 & 3.17  & $<~0.74$ & - & $<~1.79$ & - & -\\
NGC~550-3 & 3.17  & $<~0.71$ & - & $<~1.84$ & - & -\\
NGC~550-4 & $-11.17$& $<~0.53$ & - & $<~1.29$ & - & -\\
NGC~550-5 & 10.89 & $<~0.54$ & - & $<~1.40$ & - & -\\

\hline

NGC~669-0 & 0.00 & $5.90~\pm~0.78$ & $-42.366~\pm~7.6$ & $<~6.99$ & - & $<~0.30$\\
NGC~669-1 & 4.67 & $<~1.70$ & - & $<~7.55$ & - & -\\
NGC~669-2 & $-4.36$& $<~1.61$ & - & $<~8.58$ & - & -\\
NGC~669-3 & 9.88 & $3.57~\pm~0.63$ & $-215.93~\pm~2.5$ & $<~13.03$ & - & $<~0.91$\\
NGC~669-4 &$ -9.02$& $2.39~\pm~0.40$ & $238.81~\pm~10.6$ & $2.92~\pm~0.53$  & $283.79~\pm~10.5$ & $0.30~\pm~0.07$\\
NGC~669-5 & 13.93& $4.64~\pm~0.83$ & $-333.98~\pm~6.7$ & $19.00~\pm~3.91$ & $-332.49~\pm~13.1$& $1.02~\pm~0.28$\\
NGC~669-6 &$-13.62$& $4.57~\pm~0.76$ & $304.29~\pm~6.7$  & $22.96~\pm~4.34$ & $355.93~\pm~10.3$ & $1.26~\pm~0.32$\\
NGC~669-7 & 22.10& $1.26~\pm~0.18$ & $-337.68~\pm~4.9$ & $1.42~\pm~0.36$  & $-287.09~\pm~15.4$& $0.28~\pm~0.08$\\
NGC~669-8 &$-22.17$& $0.51~\pm~0.10$ & $349.13~\pm~1.3$  & $0.91~\pm~0.37$  & $427.67~\pm~5.2$  & $0.45~\pm~0.20$\\

\hline

UGC~12591-1 & 0.00 & $1.39~\pm~0.20$ & $-103.09~\pm~4.7$ & $1.99~\pm~0.38$ & $-20.10~\pm~14.0$ & $0.36~\pm~0.09$\\
UGC~12591-2 & 5.10 & $0.81~\pm~0.16$ & $-72.98~\pm~2.5$ & $<~1.51$ & - & $<~0.47$\\
UGC~12591-3 & 10.38& $1.00~\pm~0.24$ & $-81.19~\pm~11.8$ & $<~1.52$ & - & $<~0.38$\\
UGC~12591-4 & $-5.48$& $<~0.63$ & - & $<~1.49$ & - & -\\
UGC~12591-5 &$-10.86$& $<~0.87$ & - & $<~1.47$ & - & -\\
UGC~12591-6 & 20.58& $<~0.45$ & - & $<~1.01$ & - & -\\
UGC~12591-7 &$-20.96$& $<~0.45$ & - & $<~0.86$ & - & -\\

\hline
\end{tabular}\label{table:COtable}\\
\justifying\noindent
\textbf{Notes.} d is the projected distance to the galaxy's minor axis. $ I_{^{12}\mathrm{CO_{10}}}$ and $I_{^{12}\mathrm{CO_{21}}}$ are the integrated line intensities for $^{12}\mathrm{CO}~J=1-0$ and $^{12}\mathrm{CO}~J=2-1$, when detected. In cases of non-detection, a 3 $\sigma$ upper limit is used. They are both corrected for main beam and forward efficiencies. $v_{12\mathrm{CO_{10}}}$ and $v_{12\mathrm{CO_{21}}}$ are the centroid velocities derived from the intensity-weight average velocity of the Gaussian components. Regions with no detected emission lines are marked with a '-' symbol. $\rm \frac{^{12}CO_{21}}{^{12}CO_{10}} $ is the intensity line ratio between $^{12}\mathrm{CO}~J=1-0$ and  $^{12}\mathrm{CO}~J=2-1$ and corrected for beam dilution.
\end{table*}

\begin{figure*}[!ht]
    \centering
    %\textbf{ $\mathrm{UGCA~145}$} \\ 
    \includegraphics[width=0.55\textwidth]{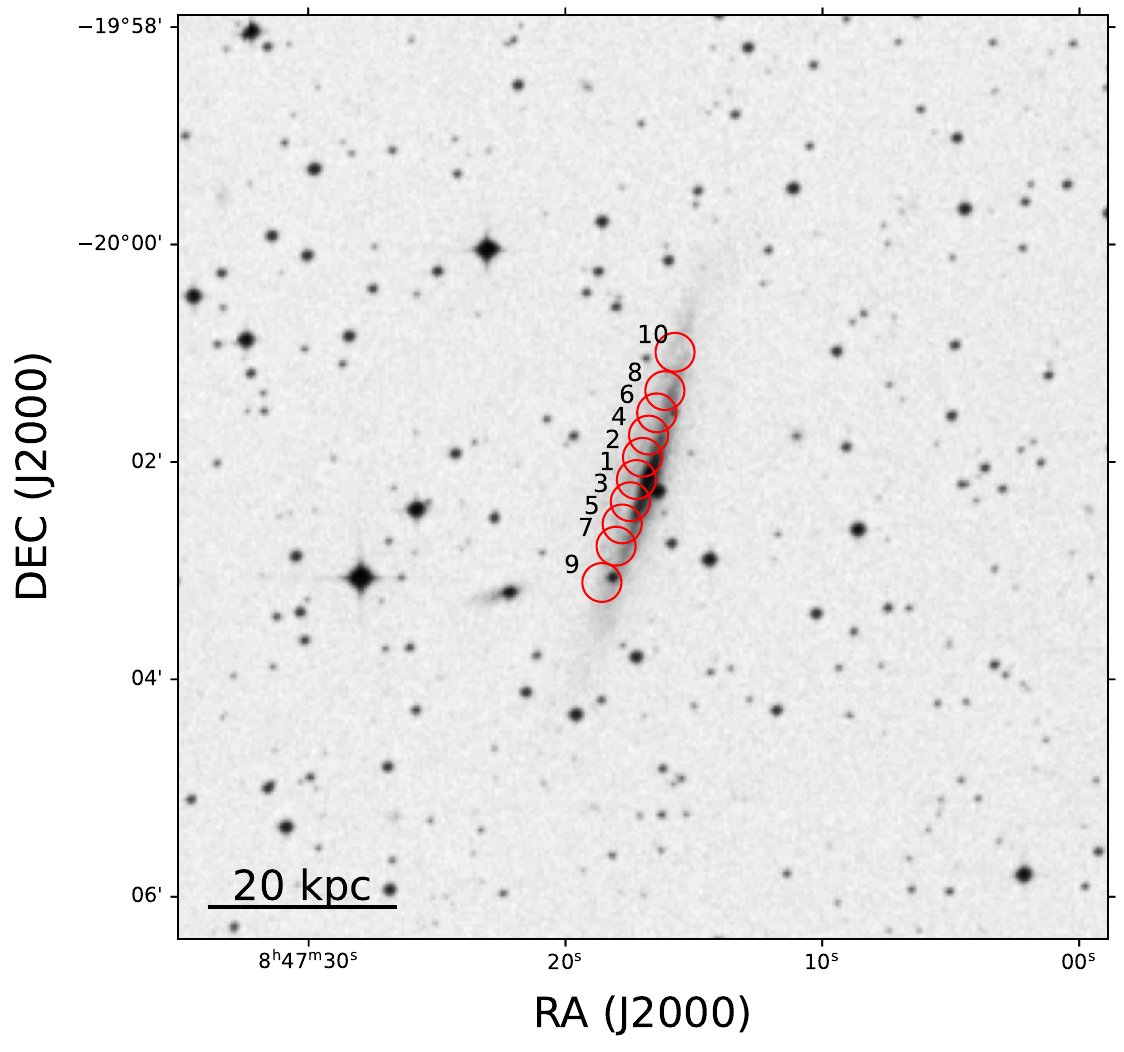}
    \hfill
    \includegraphics[width=0.44\textwidth]{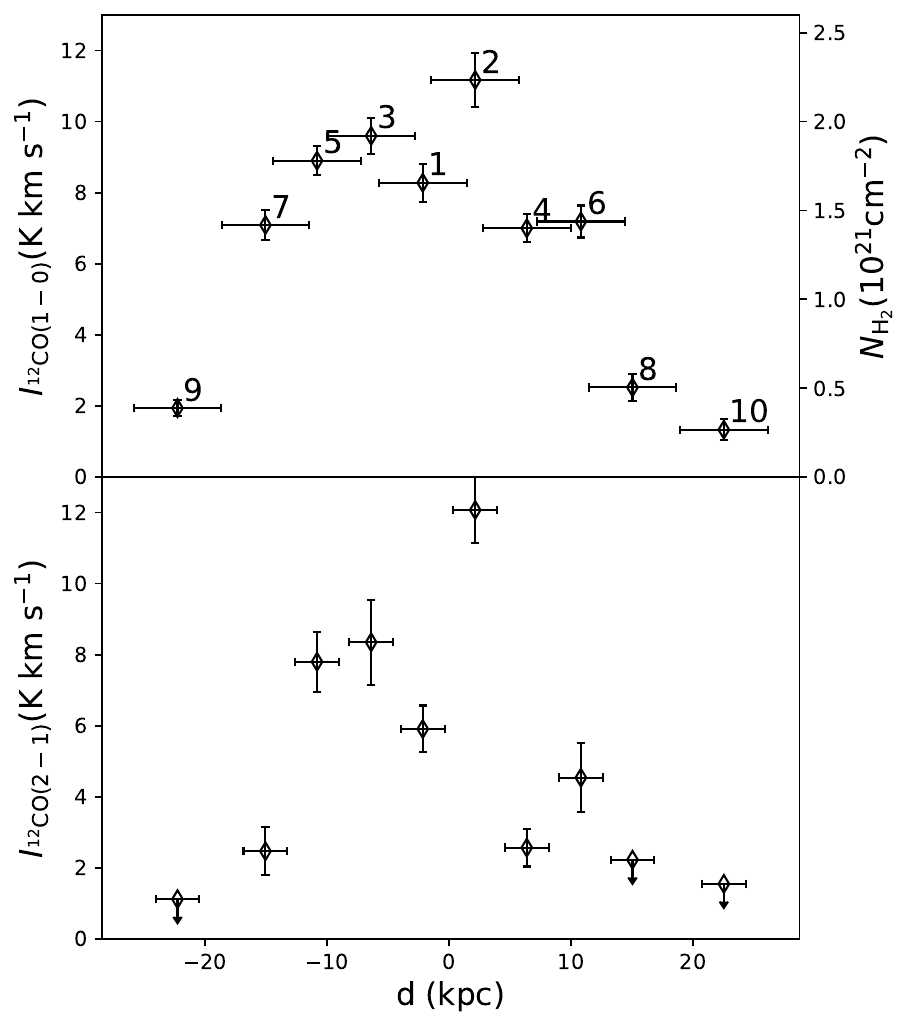}
\\
    \caption{\textit{left~panel}: DSS r-band image displays an $ 8.5^{\prime}~\times~8.5^{\prime}$ centered at UGCA~145. The solid circles are the location of IRAM 30m beams for the $^{12}\mathrm{CO}~J=1-0$ band with a diameter of $21.4^{\prime\prime}$. \textit{right~panel}: The integrated intensities of the $^{12}\mathrm{CO}~J=1-0$ (top row) and $^{12}\mathrm{CO}~J=2-1$ (bottom row) lines along the galaxy disk. The right y-axes of the top panel shows the column density of the molecular gas in these regions.}
    \label{fig:I_distance}
\end{figure*}

\section{Results}\label{sec:result}
\subsection{Total Mass of Molecular Gas}

To convert the integrated intensity of the $^{12}\mathrm{CO}~J=1\text{--}0$ emission lines into molecular hydrogen column density ($N_{\mathrm{H_2}}$), we adopt the CO-to-H$_2$ conversion factor, defined as $X_{\mathrm{CO}} = N_{\mathrm{H_2}} / I_{\mathrm{CO}}$. This factor depends on physical conditions such as metallicity, optical depth, and gas density \citetext{e.g., \citealt{glover11, wolfire10, bolatto13}}. Several methods have been employed to estimate $X_{\mathrm{CO}}$, including virial mass measurements of molecular clouds \citep{solomon87}, $\gamma$-ray observations \citep{strong96}, and modeling of diffuse X-ray emission \citep{sofue16}. These techniques yield values in normal galaxies (including star-forming dwarf, spiral, or elliptical galaxies) that cluster around a typical $X_{\mathrm{CO}}$ of $2 \times 10^{20}~\mathrm{cm^{-2}/(K~km~s^{-1})}$, consistent with the Milky Way (MW) calibration, with typical uncertainty of 0.3 dex \citep{bolatto13}. We adopt this MW value in our analysis, which would scale our derived molecular gas masses accordingly but would not affect the main conclusions of this work. The total H$_2$ mass is then calculated by multiplying the derived column density by the area of the observed regions.

%Although our galaxies are several times more massive than the Milky Way, the X-factor is found to remain broadly consistent with the Galactic value, with systematic uncertainties of about 30–50% due to metallicity and ISM conditions (e.g., Rosolowsky & Leroy 2006; Bolatto et al. 2013). Such variations would linearly scale the molecular gas masses but do not affect the relative comparisons or the main conclusions of this work.

% No matter these studies focus on larger masses, higher resolution scales, or nearby normal star-forming galaxies, the conclusions are generally consistent with those in the Milky Way, with deviations around 40 percent in extragalactic environments (e.g., Rosolowsky & Leroy 2006; Bolatto et al. 2013

Our CO line observations are obtained from discrete regions within each galaxy. To estimate the total molecular gas content of the entire galaxy, we need to rescale the directly measured CO line intensities. We use 22~$\mu$m images from the \href{https://irsa.ipac.caltech.edu/applications/wise/?__action=layout.showDropDown&}{WISE} archive \citep{WISE20} to calculate such a scaling factor for each galaxy, based on the well-established correlation between mid-infrared emission and CO line emission \citetext{e.g., \citealt{kennicutt98, Leroy08}}. For each galaxy, we subtract the local background from the WISE 22~$\mu$m image and convolve the result to match the IRAM 30m beam size at 115~GHz. We then measure the 22~$\mu$m flux within both the IRAM 30m pointings and the entire galaxy. The scaling factor, $f_{\mathrm{IR}}$, is defined as the ratio of the total galaxy flux to the flux within the IRAM 30m beams. To define the galaxy boundary, we adopt the 22~$\mu$m contour corresponding to three times the standard deviation of the background ($3\sigma$). The total molecular gas mass ($M_{\mathrm{H_2}}$) is then calculated by multiplying the directly measured molecular gas mass from the IRAM 30m beams by $f_{\mathrm{IR}}$. The resulting molecular gas masses of our sample range from $9.4 \times 10^8$ to $1.6 \times 10^{10}~\rm M_{\odot}$, and are listed in Table~\ref{table:masstable}.

\setlength{\tabcolsep}{25pt}
\begin{table*}[!ht]

\centering
\caption{Molecular gas mass of the sample galaxies} 
\footnotesize
\begin{tabular}{cccccc}
\hline\hline
Name & $f_{\mathrm{IR}}$ &$M_\mathrm{{H_2}}^o$ & $M_\mathrm{{H_2}}$ & $M_\mathrm{cold}$ & $M_\mathrm{ML}$ \\
 &  & $\mathrm{\times 10^8 M_\odot}$ &  $\mathrm{\times 10^8 M_\odot}$ & $\mathrm{\times 10^{10} M_\odot}$ & $\mathrm{\times 10^{10} M_\odot}$\\
\hline
UGCA~145 & 1.92 &$84.0^{+13.2}_{-12.8}$ & $161.3^{+25.9}_{-24.6}$ & $9.8^{+0.26}_{-0.25}$ & $1.14^{+0.01}_{-0.06}$ \\
NGC~550 & 0.99 & $<9.5$ & $<9.4$ & $<0.6$ & $1.5^{+0.02}_{-0.16}$\\
UGC~12591& 1.40 &$13.4\pm2.0$ & $18.8 \pm 2.8$ & $0.7\pm 0.03$ & $3.5^{+0.08}_{-0.44}$\\
NGC~669 & 1.52 & $42.3\pm 8.5$ & $64.3\pm 12.9$ & $1.1\pm 0.13$ & $1.7^{+0.01}_{-0.09}$\\
NGC~5908 & - & - & $83.0\pm 4.0$ & $3.6\pm 0.04$ & $1.4^{+0.01}_{-0.08}$ \\
\hline
\end{tabular}\label{table:masstable}\\
\justifying\noindent
\textbf{Notes.} $f_{\mathrm{IR}}$ is the scaling factor, derived from the intensity ratio between the selected observed regions and the area within 3 $\sigma$ contour in the \textit{WISE} $22~\rm \mu m$ galaxy image. $M_\mathrm{{H_2}}^o$ and $M_\mathrm{{H_2}}$ are the total molecular gas mass in the observed regions and the entire galaxy, respectively. $M_{\mathrm{cold}}$ is the total cold gas mass in the entire galaxy including $M_\mathrm{{H_2}}$ and $M_\mathrm{{HI}}$. $M_{\rm ML}$ is the total stellar mass loss from the galaxy accumulated over a Hubble timescale. The data of NGC~5908 is obtained from \citet{jiangtao19}.
\label{tab:masstable}
\end{table*}

\subsection{CO Line Ratios}

%The $^{12}$CO~$J=2-1$/$J=1-0$ ratio is primarily determined by the temperature and optical depth. It reflects the structure and heating sources of the molecular clouds. Gas with a low $^{12}$CO~$J=2-1$/$J=1-0$ ratio $\textless$~0.7 typically originates from the extended low-density envelopes of molecular clumps, while the high ratio in the range of 0.7-1 is also primarily influenced by optical depth \citep{Hasegawa97} and tends to arise from highly concentrated molecular clumps characterized by steep density gradients and thin CO-emitting envelopes (e.g., \citealt{Penaloza17}). For very high $^{12}$CO~$J=2-1$/$J=1-0$ ratios ($\textgreater$ 1), the local environment has a significant influence, such as UV photons from young stars, shock waves from supernova explosions \citep{Hasegawa97}, or through effective heating mechanisms in extreme starburst environments (\citealt{Papadopoulos12}), including factors like supersonic turbulence of the molecular clouds, cosmic ray ionization rates, and the interstellar radiation field (e.g., \citealt{Penaloza18}).

The $\mathrm{^{12}CO}~J=2\text{--}1/J=1\text{--}0$ line ratio is sensitive to the physical conditions of molecular clouds, such as gas temperature and optical depth, and thus provides insights into their internal structure and heating mechanisms. Based on this line ratio, Galactic molecular clouds can be broadly classified into three categories: low-ratio gas with $R_{21} < 0.7$, high-ratio gas with $0.7 \leq R_{21} \leq 1.0$, and very high-ratio gas with $R_{21} > 1.0$ \citep{hasegawa97}, where $R_{21} \equiv \mathrm{^{12}CO}~J=2\text{--}1/J=1\text{--}0$. Low ratios are generally associated with diffuse outer layers of molecular clouds, high ratios with compact clumps exhibiting steep density gradients, and very high ratios with highly excited gas produced by strong UV radiation, shocks, or extreme heating environments \citetext{e.g., \citealt{hasegawa97, penaloza17, penaloza18}}.

%The low-ratio gas is most often linked to molecular material residing in the diffuse outer layers that surround dense clumps \citep{penaloza17}. The high-ratio gas resides in compact clumps with steep density gradients, whose CO photospheres occur at higher densities \citep{hasegawa97}. The very high-ratio gas traces highly excited gas, typically driven by strong UV radiation, shocks, or enhanced heating in extreme environments \citep{penaloza17}. 

\begin{figure}[h]
    \centering
    \includegraphics[width=0.47\textwidth]{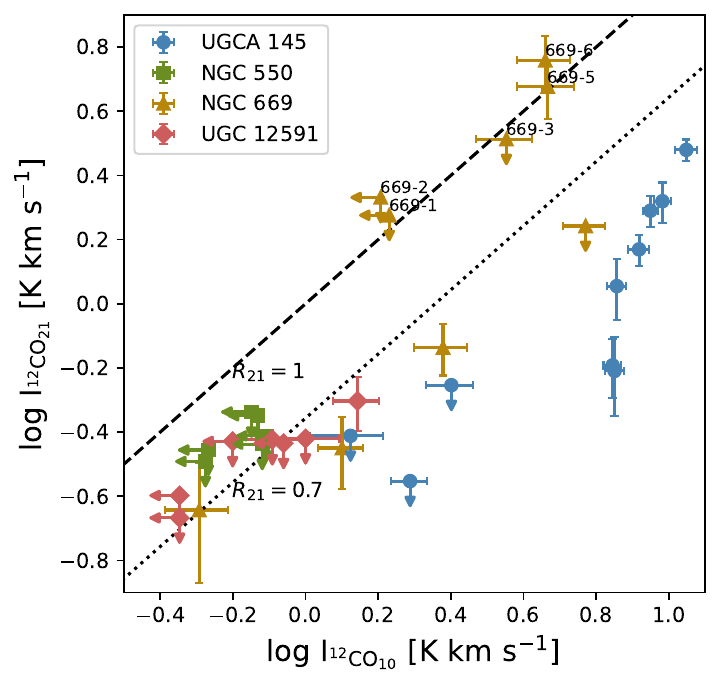}\\
    \caption{Comparison of $\mathrm{^{12}CO}~J=1\text{--}0$ and $\mathrm{^{12}CO}~J=2\text{--}1$ integrated intensities. $I_{\rm^{12}CO_{10}}$ is corrected for beam dilution. The dashed and dotted lines indicate constant intensity ratios $R_{21}=1$ and $0.7$, respectively. Different symbols denote different galaxies.}
    \label{fig:line_ratio}
\end{figure}

As shown in Figure~\ref{fig:line_ratio}, $R_{21}$ of our sample galaxies is generally below or close to 0.7 across most of the observed regions (except for NGC~669 which has a slightly higher $R_{21}$). Such ratios are typically associated with molecular gas in the extended, low-density envelopes of molecular clumps \citep{penaloza17}. The low $R_{21}$ values may indicate cooler gas temperatures and/or higher optical depths, so that external heating and disturbances cannot effectively influence the interiors of molecular clumps, resulting in a reduction in SFE. These physical conditions are consistent with the low SFE derived in \S\ref{subsec:SFE}, and are indicative of the suppressed star formation activity in these massive spiral galaxies.

%Low $R_{21}$ values may indicate cooler gas temperatures and/or higher optical depths, which can suppress star formation. Such ratios are typically associated with molecular gas in the extended, low-density envelopes of molecular clumps \citep{penaloza17}, and are indicative of relatively weak star formation activity—consistent with the known properties of these galaxies. 
%An exception is found in certain regions of NGC~669, where the line ratio exceeds 1. This elevated ratio suggests that the molecular gas in these regions may be subject to significant external heating, such as from ultraviolet (UV) radiation produced by young stars or shocks from supernovae and stellar winds \citetext{e.g., \citealt{hasegawa97}}. However, we have not yet found direct evidence to support this scenario.

%\textbf{[ljt: I deleted the discussions on NGC~669, as it does not provide anything useful. If you want to keep it, do not just speculate.]}

\section{Discussion}\label{sec:discussion}

In this section, we compare the molecular gas content with other key physical properties of the CGM-MASS galaxies. For comparison, we also include a subset of galaxies from the CHANG-ES sample (Continuum Halos in Nearby Galaxies—an EVLA Survey; \citealt{Irwin2012a,Irwin2012b}) that have spatially resolved CO observations available \citep{yan24,Yan25}.

\subsection{Molecular Gas Properties of Massive Disk Galaxies}

\subsubsection{Star Formation Efficiency} \label{subsec:SFE}
 
The star formation law, which is the relationship between the SFR surface density ($\Sigma_{\mathrm{SFR}}$) and the molecular gas surface density ($\Sigma_{\mathrm{H_2}}$), characterizes the SFE of galaxies \citetext{e.g., \citealt{kennicutt98,Kennicutt12,Leroy08}}. It provides a useful benchmark for assessing whether a galaxy is subject to some star formation suppression mechanisms that inhibit the conversion of molecular gas into stars. 

The SFR in this work is derived from the 22 $\mu$m luminosities, following the calibration of \citet{rieke09}, which assumes an IMF (initial mass function) broadly consistent with the \citet{chabrier03}. For comparison, the CHANG-ES project used a combination of 22 $\mu$m and H$\alpha$ measurements but adopted the same type of IMF. We adopt the optical isophotal diameter ($D_{25}$) as a common reference for the spatial extent of both the star-forming disk and the cold gas disk, and use it to compute $\Sigma_{\mathrm{SFR}}$ and $\Sigma_{\mathrm{H_2}}$. Considering the star-forming disk is typically smaller than the molecular gas disk, using the same area for both quantities may slightly underestimate $\Sigma_{\mathrm{SFR}}$ and overestimate $\Sigma_{\mathrm{H_2}}$ \citetext{e.g., \citealt{Leroy08, bigiel08, schruba11}}. However, this effect is typically minor and does not qualitatively impact our analysis.

\begin{figure}[h]
    \centering
    \includegraphics[width=0.47\textwidth]{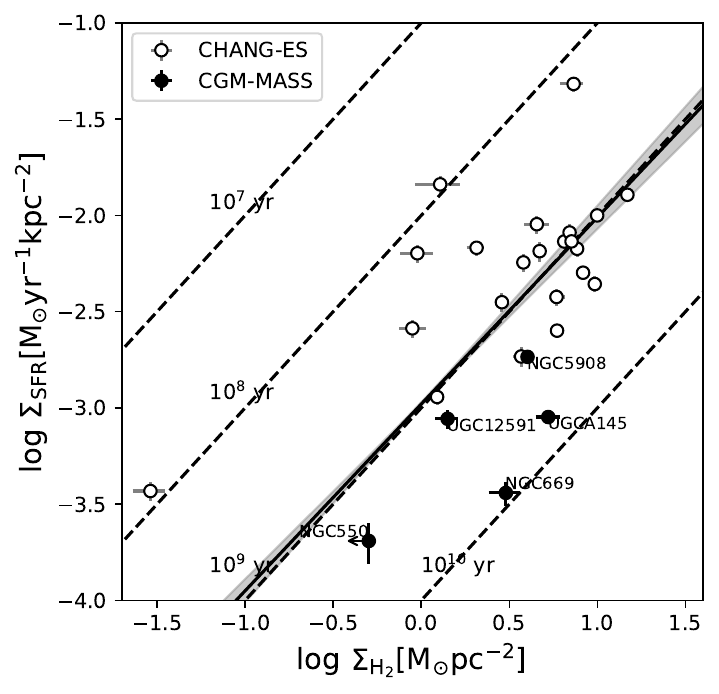}\\
    \caption{$\Sigma_{\mathrm{SFR}}$ v.s. $\Sigma_{\mathrm{H_2}}$. The solid line and the shaded area represents the best fitting of the star formation law and the 1 $\sigma$ uncertainty of the fitting provided in \citet{querejeta21}. The dashed lines from top to bottom represent gas depletion time of $10^7$, $10^8$, $10^9$, and $10^{10}$ years, respectively. }%\textbf{[ljt: I mentioned a few times, the $^{-2}$ in the x-axis should be superscript.]}}
    \label{fig:SFL}
\end{figure}

Figure~\ref{fig:SFL} presents the $\Sigma_{\mathrm{H_2}}$–$\Sigma_{\mathrm{SFR}}$ relation for our CGM-MASS galaxies, compared with galaxies from the CHANG-ES sample. The best-fit K–S law from previous studies typically corresponds to a molecular gas depletion timescale of $\tau_{\mathrm{H_2}} \equiv M_{\mathrm{H_2}}/\mathrm{SFR} \sim 2 \times 10^9~\mathrm{yr}$ \citetext{e.g., \citealt{bigiel08, querejeta21}}. For comparison, we adopt the relation from \citet{querejeta21}, which is based on spatially resolved, sub-galactic measurements of both $\Sigma_{\mathrm{H_2}}$ and $\Sigma_{\mathrm{SFR}}$. As shown in Figure~\ref{fig:SFL}, the SFE of the CHANG-ES galaxies generally follows the best-fit K–S relation. In contrast, the CGM-MASS galaxies exhibit systematically lower SFEs, particularly NGC~669 and UGCA~145, whose gas depletion timescales are nearly an order of magnitude longer than the canonical value. This suggests the presence of effective star formation quenching mechanisms in these most massive, isolated spiral galaxies.

We further explore which quenching mechanisms may be responsible for suppressing star formation in the CGM-MASS galaxies. As discussed in \S3.2, the low $R_{21}$ values indicate physical conditions that may suppress star formation by insulating molecular clumps from external heating and disturbances. In addition to this, a variety of physical processes can also inhibit star formation, including supernova feedback \citep{kay02, hopkins14, marri03}, AGN feedback \citep{piotrowska22, maiolino12}, ram-pressure stripping \citep{boselli22, abadi99}, virial shock heating \citep{birnboim03, dekel06}, and morphological quenching \citep{martig09}. Supernova feedback is particularly effective at regulating star formation in low-mass galaxies, but it generally has limited impact in more massive systems such as those in the CGM-MASS sample. AGN feedback, another strong quenching candidate in massive galaxies, is unlikely to be a dominant factor here due to the lack of observational evidence for strong ongoing AGN activity in these galaxies \citep{jiangtao17}. Ram-pressure stripping, which requires a dense ambient medium and typically occurs in galaxy clusters, is also unlikely to be important given the isolated environments of the CGM-MASS galaxies \citep{boselli22}. Virial shock heating is expected in massive halos, where infalling gas is shock-heated to the virial temperature near the halo boundary \citep{birnboim03, dekel06}. While this mechanism may have been important at high redshift, when the gas accretion rate was higher \citep{nelson15, vandeVoort11}, there is no strong evidence that it plays a major role in quenching star formation in similar massive galaxies in the local universe. 

Despite the various mechanisms discussed above, most of them seem unlikely to play a dominant role in these galaxies. By contrast, morphological quenching occurs when a dynamically hot stellar component, such as a massive spheroid, stabilizes the gas disk against gravitational collapse, thereby inhibiting star formation despite the presence of cold gas \citep{martig09}. 
The morphology types of CGM-MASS galaxies are generally earlier than Sb (with the exception of UGCA 145, which is classified as Sbc), mostly corresponding to early-type spirals that are generally characterized by prominent bulges.
The presence of prominent bulges in the CGM-MASS galaxies suggests that morphological quenching may play a significant role in suppressing their star formation, as also been suggested in some S0 galaxies (e.g., \citealt{Li2009,Li2011}). Overall, morphological quenching likely contributes to the suppression of star formation in these galaxies, but the detailed interplay of processes remains complex and not yet fully understood.

\subsubsection{Cold Gas Budget}

\begin{figure}[h]
    \centering
    \includegraphics[width=0.47\textwidth]{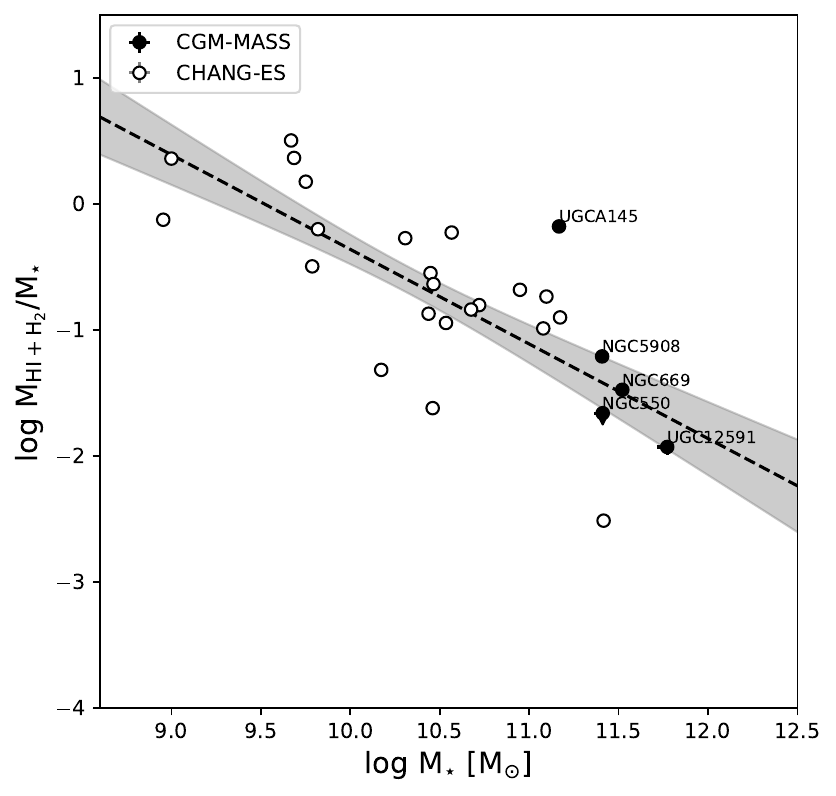}\\
    \caption{The distribution of $M_{\mathrm{HI+H_2}}$ with $M_{\star}$. The dashed line and shaded area represent the best-fitting linear relation and the 1$\sigma$ uncertainty of the fitting using the CHANG-ES data from \citet{Yan25}.}
    \label{fig:H2-star}
\end{figure}

The ratio of total cold gas mass to stellar mass ($M_{\mathrm{HI+H_2}} / M_{\star}$) and the ratio of molecular to atomic gas mass ($M_{\mathrm{H_2}} / M_{\mathrm{HI}}$) are often more informative metrics than the absolute mass or surface density of molecular gas when assessing a galaxy's gas richness. As shown in Figure~\ref{fig:H2-star}, the CHANG-ES data reveal a weak negative correlation between $M_{\mathrm{H_2+HI}} / M_{\star}$ and stellar mass. We also plot $M_{\mathrm{H_2+HI}} / M_{\star}$ as a function of stellar mass for the CGM-MASS samples and compare the results with the relation fitted by CHANG-ES data. All samples generally follow the declining trend, with more massive galaxies having lower $M_{\mathrm{H_2+HI}} / M_{\star}$.

\begin{figure}[h]
    \centering
    \includegraphics[width=0.47\textwidth]{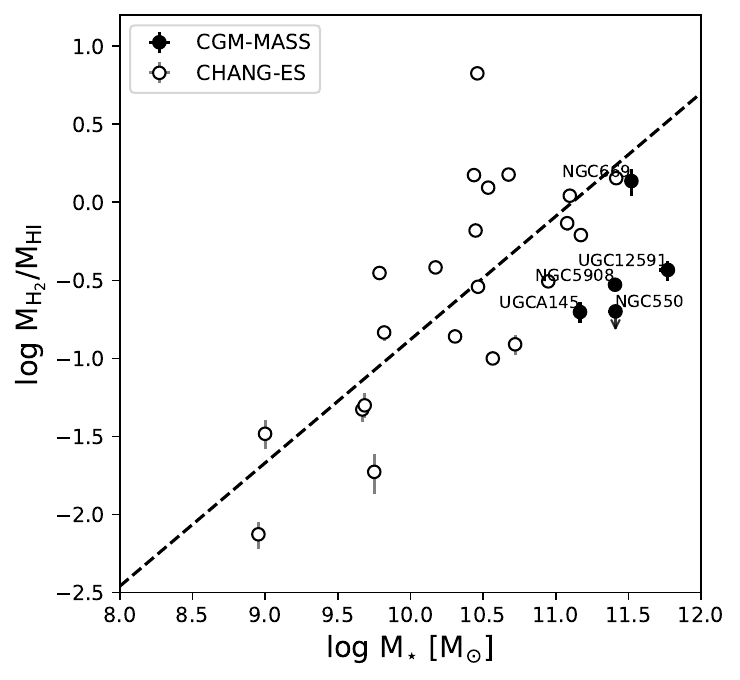}\\
    \caption{The distribution of $M_{\mathrm{H_2}}/M_{\mathrm{HI}}$ ratio with $M_{\star}$. The dashed line represents the result fitted using the CHANG-ES data from \citet{Yan25}.}
    \label{fig:H2-HI}
\end{figure}

The mass ratio of molecular to atomic gas $M_{\mathrm{H_2}} / M_{\mathrm{HI}}$ serves as a diagnostic of the efficiency with which atomic gas is converted into molecular form. This conversion efficiency is primarily influenced by factors such as metallicity and interstellar pressure, with higher metallicities and greater pressures favoring the formation of molecular gas \citep{elmegreen1993h, Blitz06,Krumholz2009}. In galaxies, both metallicity and mid-plane pressure tend to increase with stellar mass—metallicity through the mass–metallicity relation, and pressure due to deeper gravitational potential wells. As a result, the $M_{\mathrm{H_2}} / M_{\mathrm{HI}}$ generally increases with stellar mass (Figure~\ref{fig:H2-HI}), consistent with trends found in previous studies \citetext{e.g., \citealt{saintonge11, Yan25}}.

Due to the lack of spatially resolved \ion{H}{1} maps for our sample, we directly compare the total atomic gas mass and molecular gas mass within the galactic disks. Figure~\ref{fig:H2-HI} shows the $M_{\mathrm{H_2}}/M_{\mathrm{HI}}$ as a function of stellar mass, alongside the best-fit relation derived from the CHANG-ES sample \citep{Yan25}. The CGM-MASS galaxies systematically lie below this relation, indicating that they are relatively deficient in molecular gas compared to atomic gas for their stellar mass. This molecular deficiency may reflect inefficiencies in the conversion of atomic to molecular gas in massive spiral galaxies. Although high stellar mass is generally associated with conditions favorable for molecular gas formation (e.g., high metallicity and mid-plane pressure), other factors may counteract this trend in these galaxies (e.g., \citealt{Krumholz2009}). For example, heating of the interstellar medium by an evolved stellar population \citep{Leroy08} or turbulent feedback from past starbursts \citep{Hayward2017} could prevent atomic gas from cooling and forming molecules. Additionally, a lower cold gas accretion rate in the isolated environments of CGM-MASS galaxies may reduce the overall supply of fresh gas available for molecular cloud formation (e.g., \citealt{Feldmann2015}). It is worth noting that new high-resolution deep \ion{H}{1} observations of NGC~5908 suggest a significantly larger atomic gas mass of $7.5 \times 10^9~\mathrm{M_\odot}$ adding contributions from an extended \ion{H}{1} envelope (Yang et al., in preparation), compared to the value \citet{jiangtao19} cited from \citet{springob05}. We anticipate that future similarly high-quality \ion{H}{1} observations will help clarify the molecular-to-atomic gas balance in these massive spiral galaxies.

\subsubsection{Origin of Cold Gas in Massive Spirals}

The supply of cold gas in galaxies can originate from various mechanisms, including accretion from the circumgalactic or intergalactic medium (CGM or IGM; in either cold or hot modes), stellar mass loss from evolved stars, and residual cold gas left over from the galaxy formation epoch \citetext{e.g., \citealt{kerevs05, leitner11, irwin95, jiangtao19}}.

We first consider stellar mass loss as a potential internal source. Using the relation provided by \citet{Li2009}, the stellar mass loss rate through evolved planetary nebulae (major form for low-mass stars) is estimated as $3.3 \times 10^{-12}~\mathrm{M_\odot~yr^{-1}~L_\odot^{-1}}$ in the K-band. By combining this mass loss rate with the stellar mass of the galaxies, we calculate the total mass return over a 10~Gyr timescale for each CGM-MASS galaxy. These values are listed in Table~\ref{table:masstable}. In most cases, the cold gas mass is comparable to or exceeds the cumulative stellar mass loss, expect for UGC~12591, where the returned cold gas from stellar mass loss may be marginally sufficient to explain the current molecular gas content. Furthermore, according to simulations by \citet{parriott08}, only a small fraction (typically $\lesssim 25\%$) of the mass lost by stars is expected to remain in the cold phase, as much of it is quickly mixed with the ambient hot gas or heated by either the gravitational potential or old stellar feedback. Therefore, we conclude that stellar mass loss alone appears insufficient to account for the observed small cold gas content.

\begin{figure}[h]
    \centering
    \includegraphics[width=0.47\textwidth]{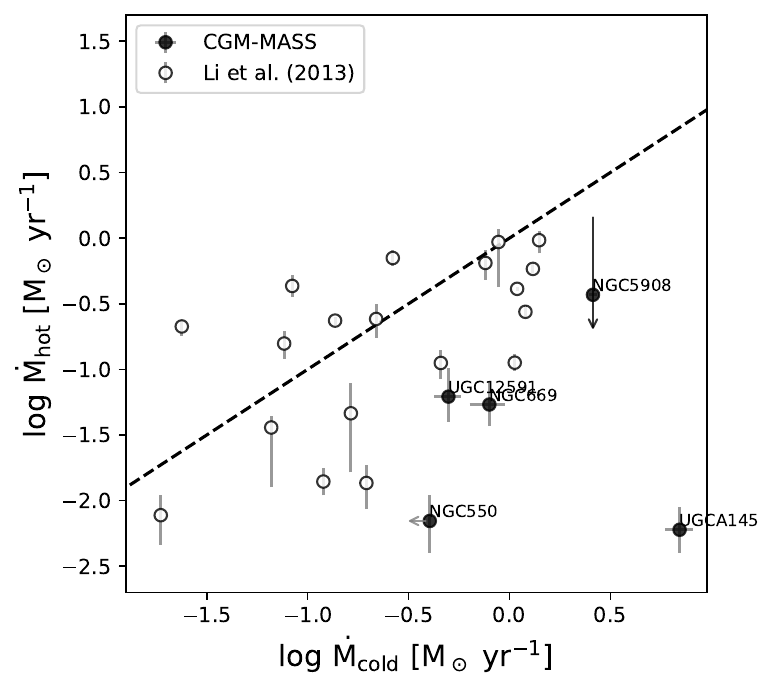}\\
    \caption{Comparison between the hot gas cooling rate and the cold gas accumulation rate. The dashed line shows where the two rates are equal ($\dot{M}_{\rm{hot}} = \dot{M}_\mathrm{cold}$).}
    \label{fig:hot_cold}
\end{figure}

We next consider accretion from the CGM. In low-mass galaxies (baryonic mass $M_{\mathrm{gal}} \lesssim 10^{10.3}~\mathrm{M_\odot}$), cold-mode accretion is thought to dominate, with gas falling directly into the disk without efficient gravitational shock heating to the X-ray emitting temperature \citep{kerevs05, nelson13}. In contrast, more massive halos like those of the CGM-MASS galaxies ($M_{\mathrm{gal}} \gtrsim 10^{11}~\mathrm{M_\odot}$) are expected to accrete gas primarily via the hot mode, in which infalling gas is shock-heated to the virial temperature and must cool radiatively before forming stars or molecular clouds. To evaluate the viability of hot-mode accretion, we use the radiative cooling rates ($\dot{M}_{\mathrm{hot}}$) derived in \citet{jiangtao17}, measured within the cooling radius defined as the radius where the cooling time equals 10~Gyr. We estimate the average cold gas accumulation rates as the total cold gas mass divided over a similar timescale of 10~Gyr and compare these rates to the radiative cooling rates to assess whether hot-mode accretion can sustain the molecular gas content. For comparison, we also include galaxies from the Chandra survey of nearby edge-on galaxies \citep{jiangtao13}. As shown in Figure~\ref{fig:hot_cold}, $\dot{M}_{\mathrm{hot}}$ of the reference galaxies from \citet{jiangtao13} is roughly comparable to the cold gas accumulation rate. In contrast, the CGM-MASS galaxies fall systematically below the line where the two rates are comparable to each other, so the hot-mode accretion rates are significantly too low to replenish their current cold gas reservoirs. 
%(\textbf{[ljt: are you using the same timescale? Please explain how did you compute the cold gas accumulation rate]}) 

For NGC~5908, there may be an additional supply of cold gas through interaction with its companion galaxy NGC~5905. Recent observations suggest the presence of ongoing \ion{H}{1} gas transfer between the two galaxies (Yang et al., in preparation). In contrast, the other four CGM-MASS galaxies appear to reside in truly isolated environments, with no signs of recent interactions or nearby companions.

As discussed above, neither stellar mass loss nor hot-mode accretion is sufficient to account for the observed molecular gas content in most of the CGM-MASS galaxies. Therefore, additional sources of cold gas must be considered. One possibility is that these galaxies experienced significant cold gas accretion in the past, such as from major mergers, gas-rich minor mergers, or filamentary cold flows that penetrated the hot halos \citep{keres09, stewart11, vandeVoort11}. In this scenario, the current molecular gas reservoirs may be long-lived remnants from earlier evolutionary phases, rather than being sustained by ongoing gas supply in the present-day isolated environment. It should be emphasized that the isolation of these galaxies indicates that any merger activity must have occurred far back in cosmic history, with companions fully accreted and tidal signatures (e.g., tidal tails) erased by dynamical processes. Consequently, the systems we observe today are quiescent disks that have long since settled back into equilibrium. This interpretation implies that the cold gas content in massive spiral galaxies may retain memory of their past assembly histories, and that present-day isolation does not necessarily imply a passive gas supply history.

%The isolated nature of these galaxies suggests that any past merger events must have occurred sufficiently early in the cosmological past, such that companion galaxies have been completely accreted, and tidal features (e.g., tidal tails) have subsequently faded due to dynamical friction. This implies that what we are observing are quiescent disks that have long since re-established equilibrium following the mergers.

\subsection{Baryonic Tully-Fisher Relation}

The Tully-Fisher relation is an empirical correlation between the rotation velocity and luminosity of spiral galaxies, expressed as $L \propto V_{\mathrm{rot}}^n$ \citep{tully77}. This relation was later generalized to the baryonic Tully-Fisher relation (BTFR), which accounts for both stellar and gas mass. \citet{mcgaugh05} confirmed the BTFR and derived a best-fit relation of $M_{\mathrm{b}} = 50V_{f}^4$, where $M_{\mathrm{b}}$ represents the total measured baryonic mass, and $V_f$ is the flat rotation velocity, showing excellent consistency for gas-rich, low-mass galaxies. However, more massive systems ($M_{\star} > 2 \times 10^{11}~\mathrm{M_{\odot}}$) often deviate from this linear relation, possibly due to factors such as angular momentum loss, a higher dark matter fraction, or baryon redistribution \citep{noordermeer07, Ogle19}.

\begin{figure}[h]
    \centering
    \includegraphics[width=0.47\textwidth]{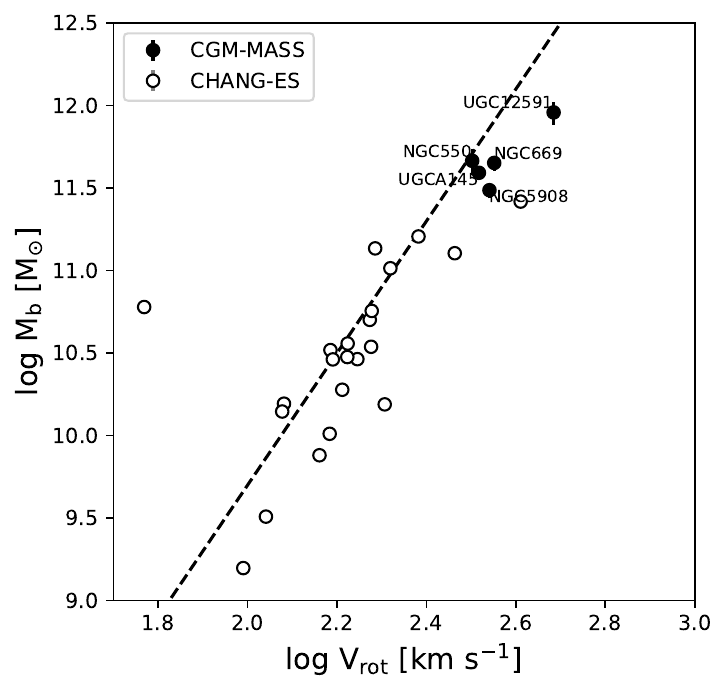}\\
    \caption{The relation between $M_{\mathrm{b}}$ and $V_{\mathrm{rot}}$. The baryonic mass of the CGM-MASS is the sum of the stellar mass, hot gas mass, and cold gas mass. The dashed line is the well defined baryonic Tully-Fisher relation from \citet{mcgaugh05}.}
    \label{fig:TF}
\end{figure}

In the CGM-MASS galaxies, the molecular gas mass measured in this work contributes less than 10\% of the total baryon budget. Nevertheless, we include a BTFR analysis here, adding the molecular gas for completeness. Following \citet{jiangtao17}, we define the total baryonic mass as the sum of stellar mass, hot gas mass, and cold gas mass ($M_{\mathrm{H_2}} + M_{\mathrm{HI}}$). We directly adopt the inclination corrected maximum gas rotation $V_\mathrm{rot}$ as a proxy for $V_f$, noting that the two values are often comparable, and A similar substitution was also adopted in the CHANG-ES sample \citep{Yan25}. We then plot the $M_{\mathrm{b}}$–$V_{\mathrm{rot}}$ relation in Figure~\ref{fig:TF} and compare our sample with the canonical BTFR from \citet{mcgaugh05}.

The inclusion of cold gas does not compensate for the previously noted baryonic mass deficit in these massive systems—our CGM-MASS galaxies still fall systematically below the best-fit BTFR. This discrepancy raises two possibilities. First, there may be a physical turnover or flattening of the BTFR at the high-mass end. Such a trend is qualitatively consistent with predictions from abundance matching models of the stellar-to-halo mass (SHM) relation, where the efficiency of converting baryons into stars declines sharply in halos above $\sim10^{12}~\mathrm{M_{\odot}}$ \citep{Behroozi2013}. Second, the observed offset may arise from missing baryonic components in the current budget.

Several studies suggest that a significant fraction of baryons in massive galaxies may reside beyond the virial radius. \citet{Li2018,Bregman18,Bregman22} showed that the baryon-to-dark matter ratio approaches the cosmic mean only when gas at large radii ($r > r_{200}$) is included. Furthermore, warm and cool gas phases traced by UV lines may contribute non-negligibly to the baryon budget \citep{werk14}, but are not available for massive spiral galaxies. A more complete census of baryonic components—particularly in hot, warm, and diffuse phases—will be essential to fully characterize the BTFR at the massive end (e.g., \citealt{Li2020}).

\section{Summary and Conclusion}\label{sec:conclusion}

We have conducted IRAM 30m observations of the $^{12}\mathrm{CO}~J=1\text{--}0$ and $^{12}\mathrm{CO}~J=2\text{--}1$ emission lines in five massive spiral galaxies selected from the CGM-MASS sample. After data reduction, we obtained integrated line intensities or $3\sigma$ upper limits for each pointing. By combining the CO measurements with \textit{WISE} 22~$\mu$m imaging and applying a beam-scaling correction, we derived the total molecular gas masses for these galaxies. We then examined their molecular gas properties, star formation efficiencies, and gas origins, and briefly assessed their positions on the baryonic Tully-Fisher relation (BTFR). Our main conclusions are summarized below:

\begin{itemize}
\item The star formation efficiencies (SFEs) in the CGM-MASS galaxies are systematically lower than the expectation from the canonical Kennicutt–Schmidt relation. This indicates that star formation is being suppressed despite the presence of molecular gas. Among various quenching mechanisms considered, morphological quenching—via the stabilizing effect of massive stellar bulges—emerges as a plausible explanation.

\item The total cold gas fraction ($M_{\mathrm{H_2+HI}} / M_\star$) declines with increasing stellar mass, in agreement with previous studies. However, the molecular-to-atomic gas mass ratios ($M_{\mathrm{H_2}} / M_{\mathrm{HI}}$) are systematically lower than expected. While this may reflect inefficiencies in atomic-to-molecular gas conversion, it could also result from uncertainties in atomic gas measurements.

\item Several potential sources of cold gas—such as stellar mass loss and hot-mode accretion—are found to be insufficient to account for the observed molecular gas content. We conclude that the molecular gas is likely a remnant from past starbursts or gas-rich mergers, rather than being sustained by current gas accretion in the present isolated environment.

\item Although the newly measured molecular gas masses are included in the baryon budget, the CGM-MASS galaxies still fall below the best-fit BTFR. This residual offset may reflect either a turnover in the BTFR at the high-mass end or the presence of additional, unaccounted baryonic components such as warm or diffuse hot gas beyond the virial radius.
\end{itemize}

\begin{acknowledgments}
J.T.L. acknowledges the financial support from the China Manned Space Program with grant no. CMS-CSST-2025-A10 and CMS-CSST-2025-A04, and the National Science Foundation of China (NSFC) through the grants 12321003 and 12273111.

\end{acknowledgments}

\renewcommand{\thefigure}{A.\arabic{figure}}
\setcounter{figure}{0}

\appendix

\section{Observation Regions and The Radial CO Lines Profile }\label{App1}

\begin{figure*}[!ht]
    \centering
    \textbf{ $\mathrm{NGC~550}$} \\ 
    \includegraphics[width=0.55\textwidth]{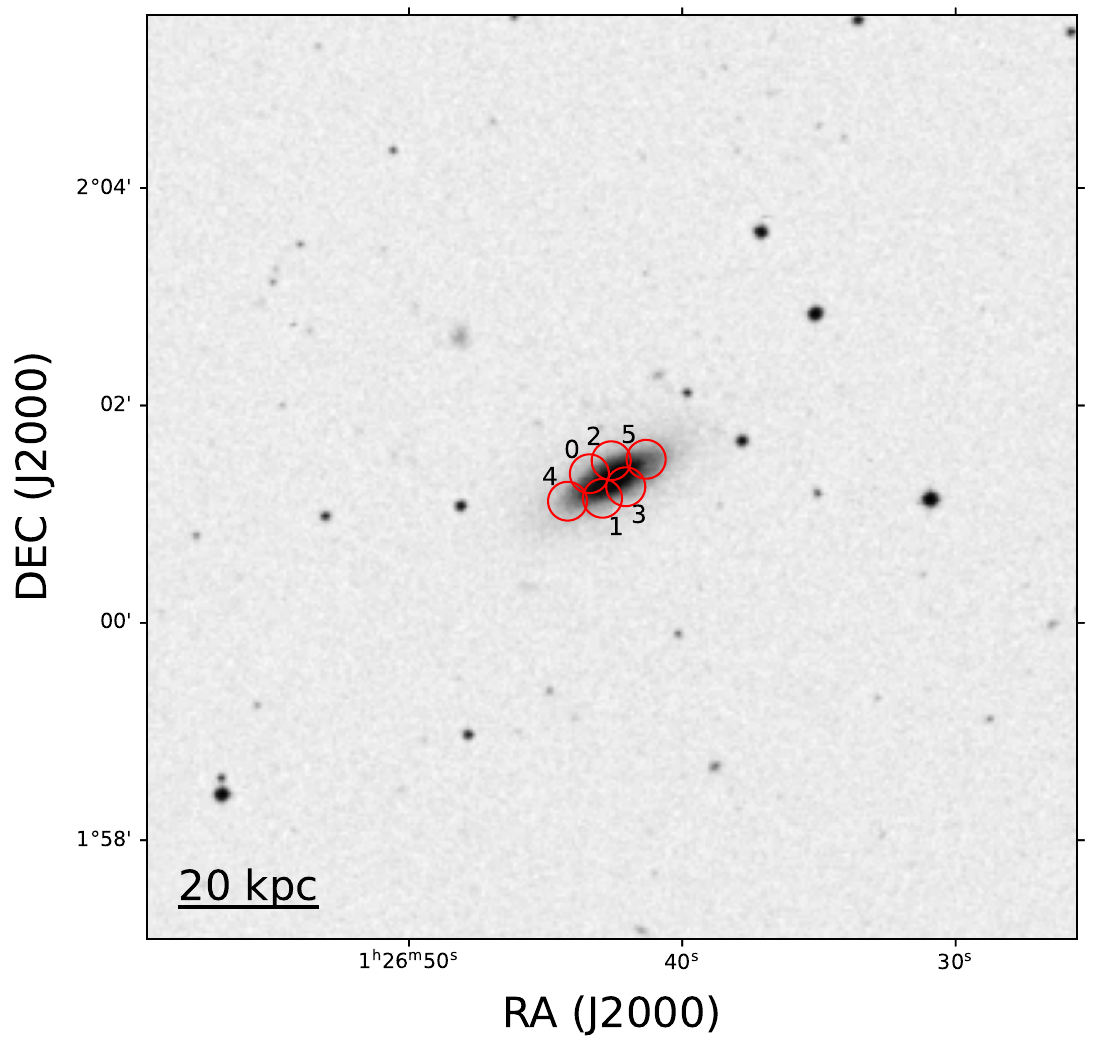}
    \hfill
    \includegraphics[width=0.44\textwidth]{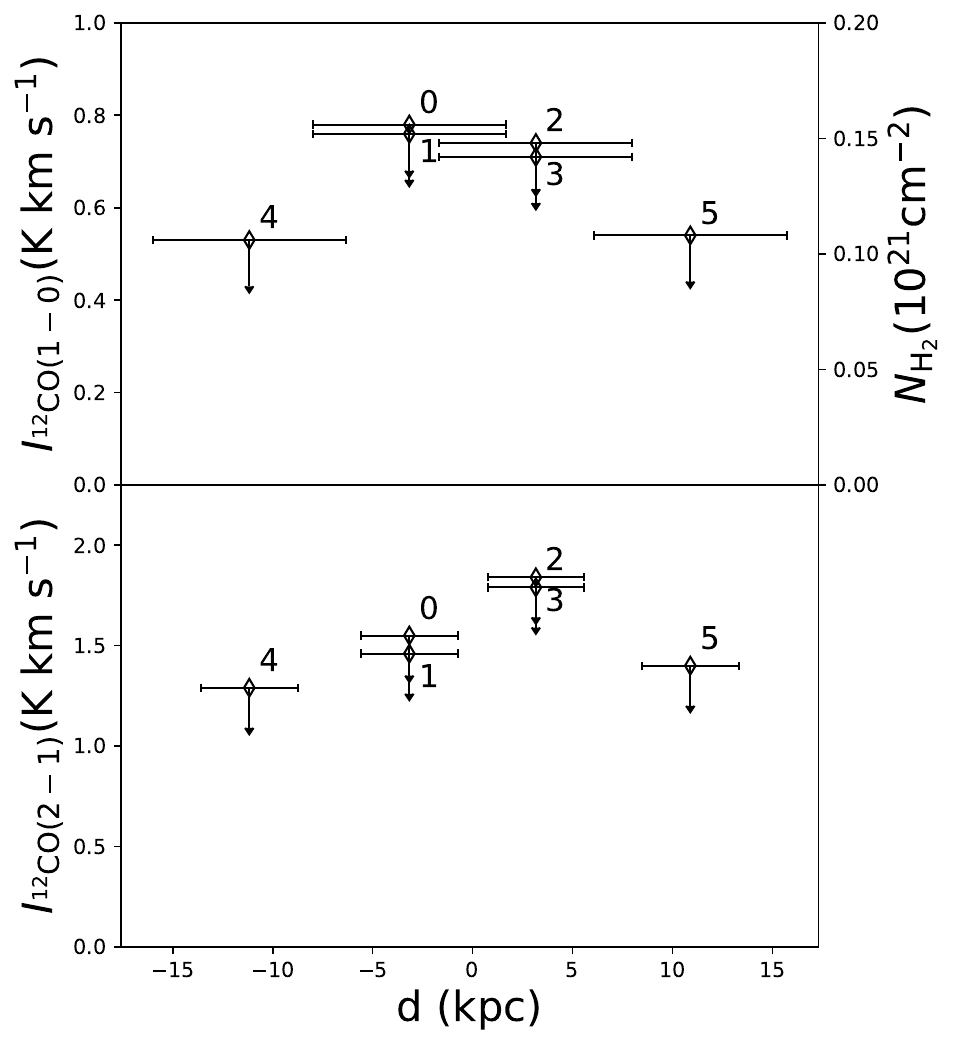}
\\
    \caption{The observation positions and the radial CO lines profile of NGC~550. Similar as Figure~\ref{fig:I_distance}.}
\end{figure*}

\begin{figure*}[!ht]
    \centering
    \textbf{ $\mathrm{NGC~669}$} \\ 
    \includegraphics[width=0.55\textwidth]{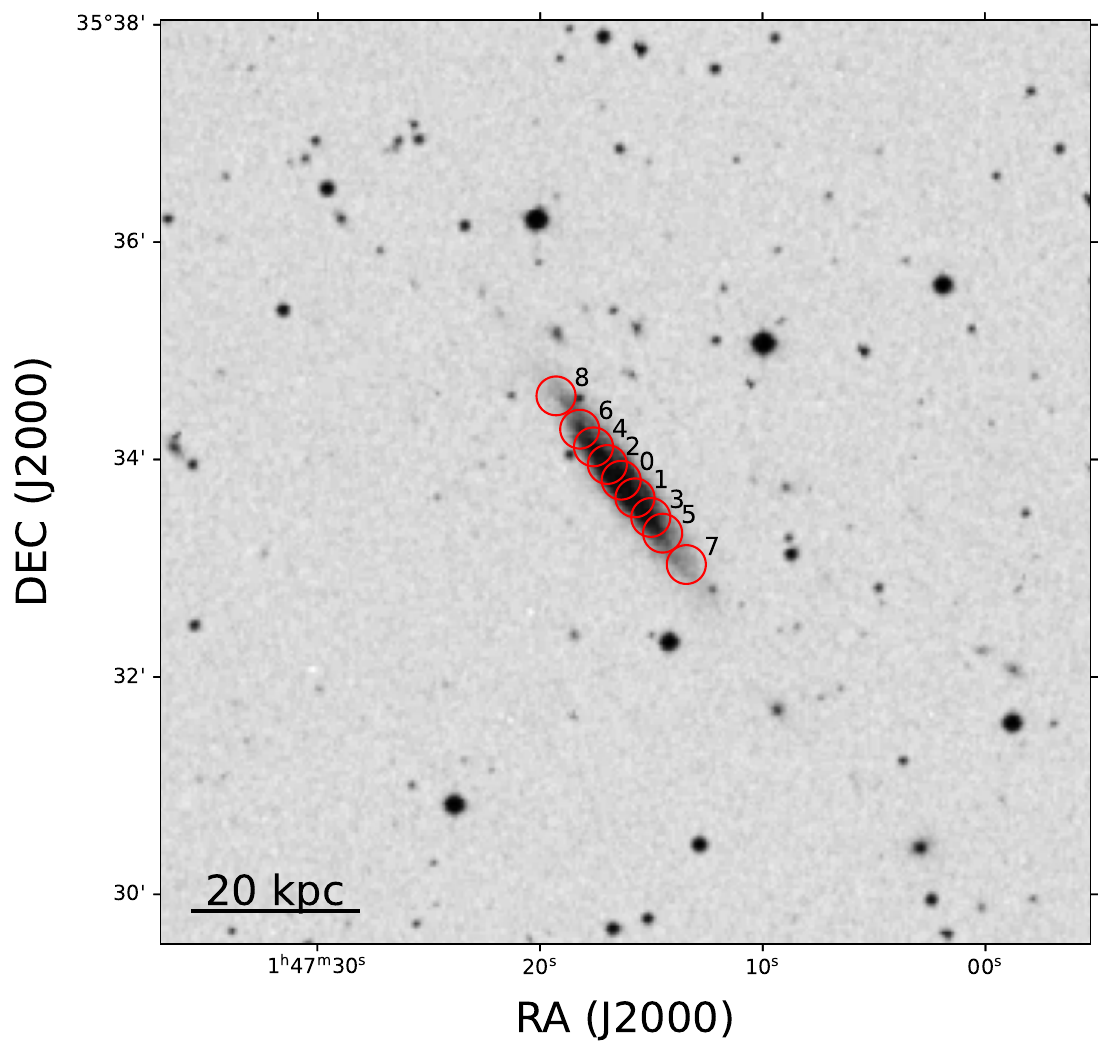}
    \hfill
    \includegraphics[width=0.44\textwidth]{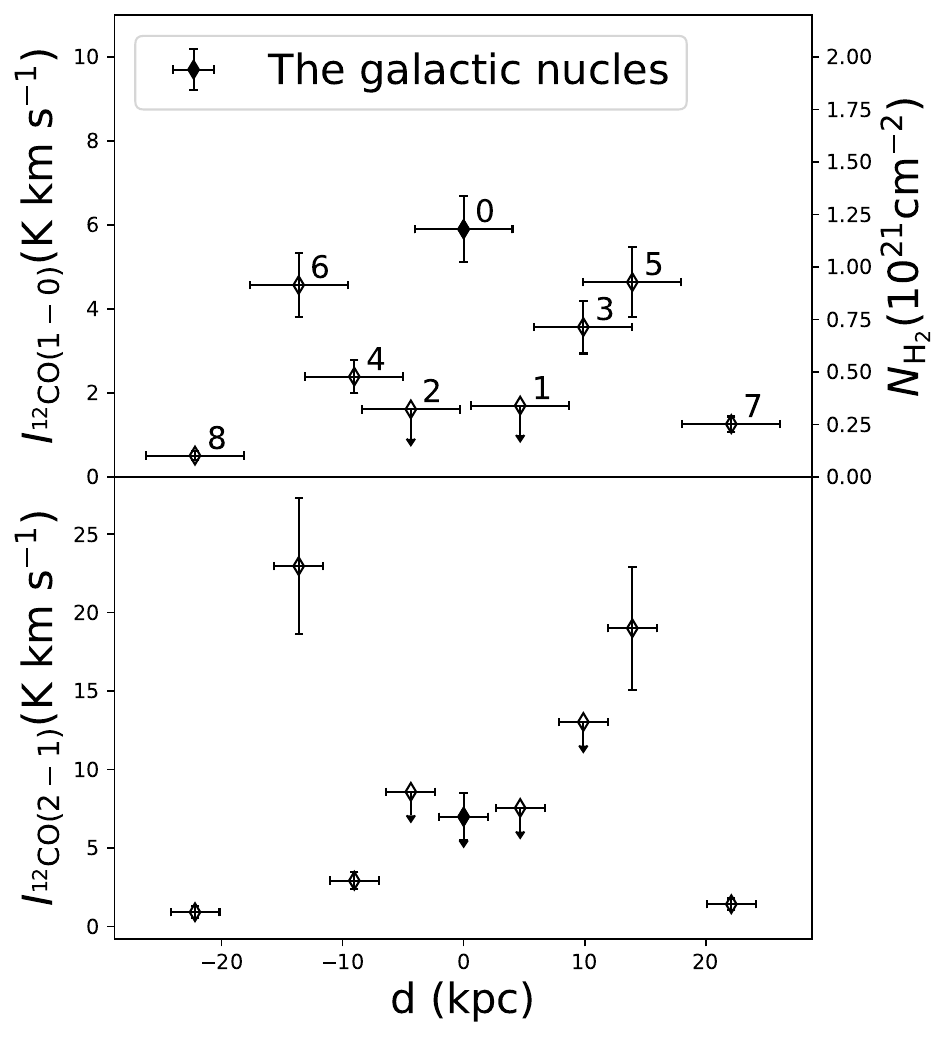}
\\
    \caption{The observation positions and the radial CO lines profile of NGC~669. Similar as Figure~\ref{fig:I_distance}.}
\end{figure*}

\begin{figure*}[!ht]
    \centering
    \textbf{ $\mathrm{UGC12591}$} \\ 
    \includegraphics[width=0.55\textwidth]{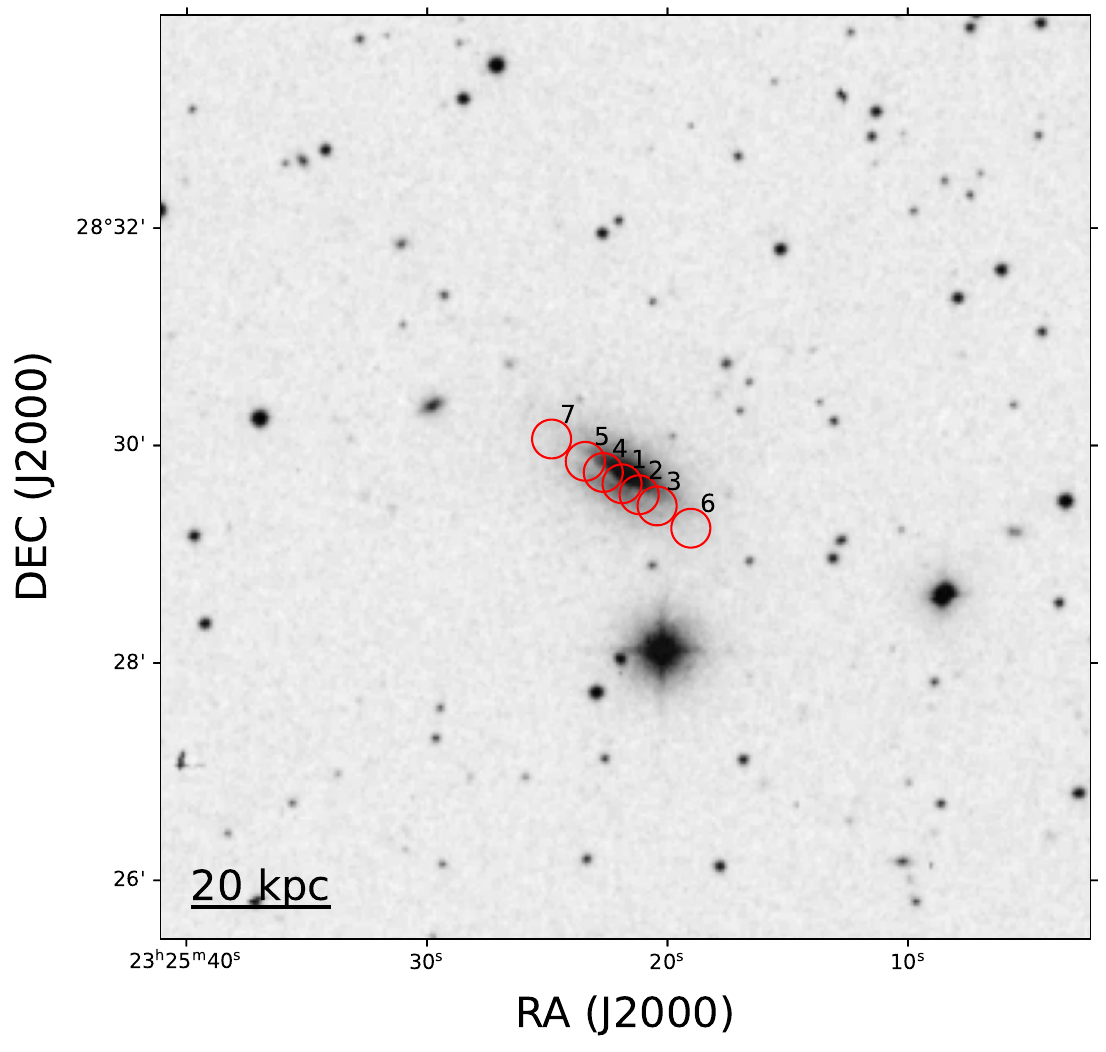}
    \hfill
    \includegraphics[width=0.44\textwidth]{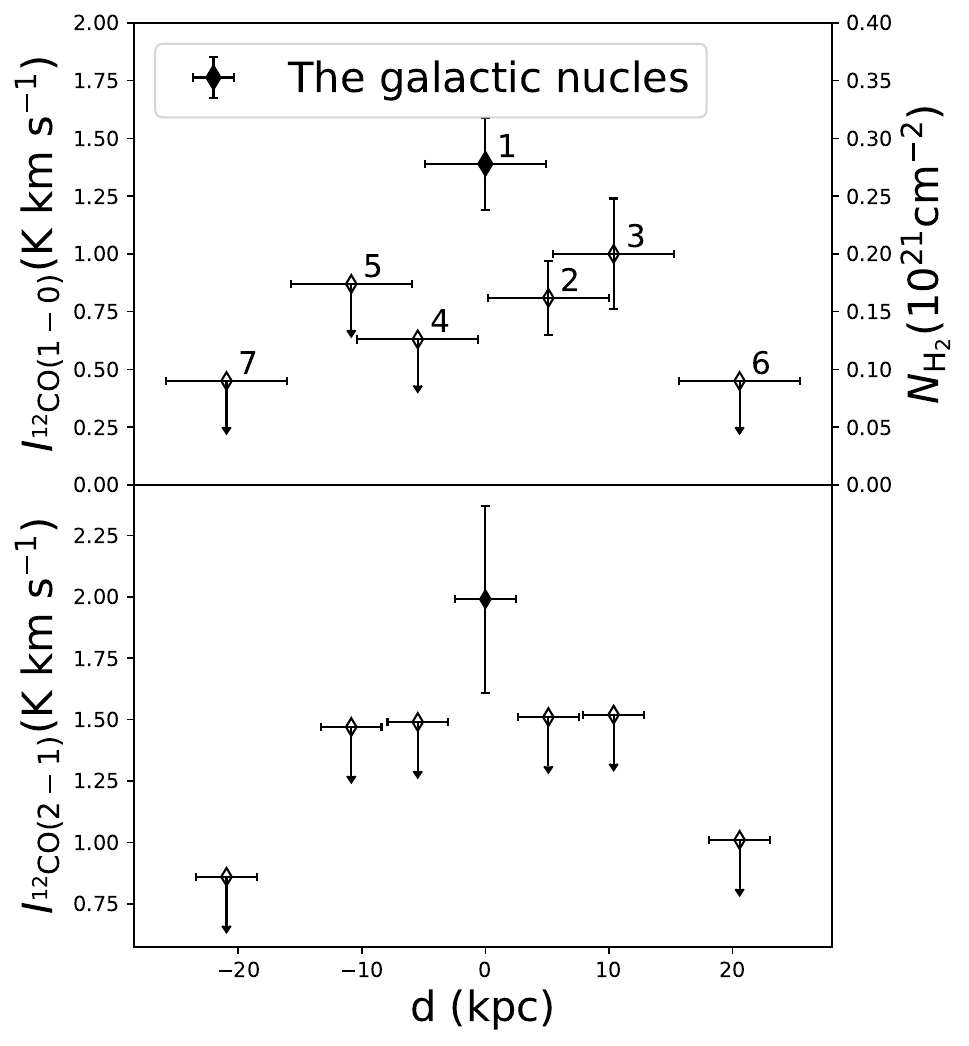}
\\
    \caption{The observation positions and the radial CO lines profile of UGC~12591. Similar as Figure~\ref{fig:I_distance}.}
\end{figure*}
\clearpage

\renewcommand{\thefigure}{B.\arabic{figure}}
\setcounter{figure}{0}

\section{CO Line Spectra}\label{App2}

\begin{figure*}[!ht]
    \centering
    \textbf{ $\mathrm{Frequency~(Hz)}$} \\
    \raisebox{6cm}{\rotatebox{90}{\centering \textbf{T (K)}}}\hspace{1em}
    \includegraphics[viewport=0 2050 650 4100,clip,width=0.22\textwidth]{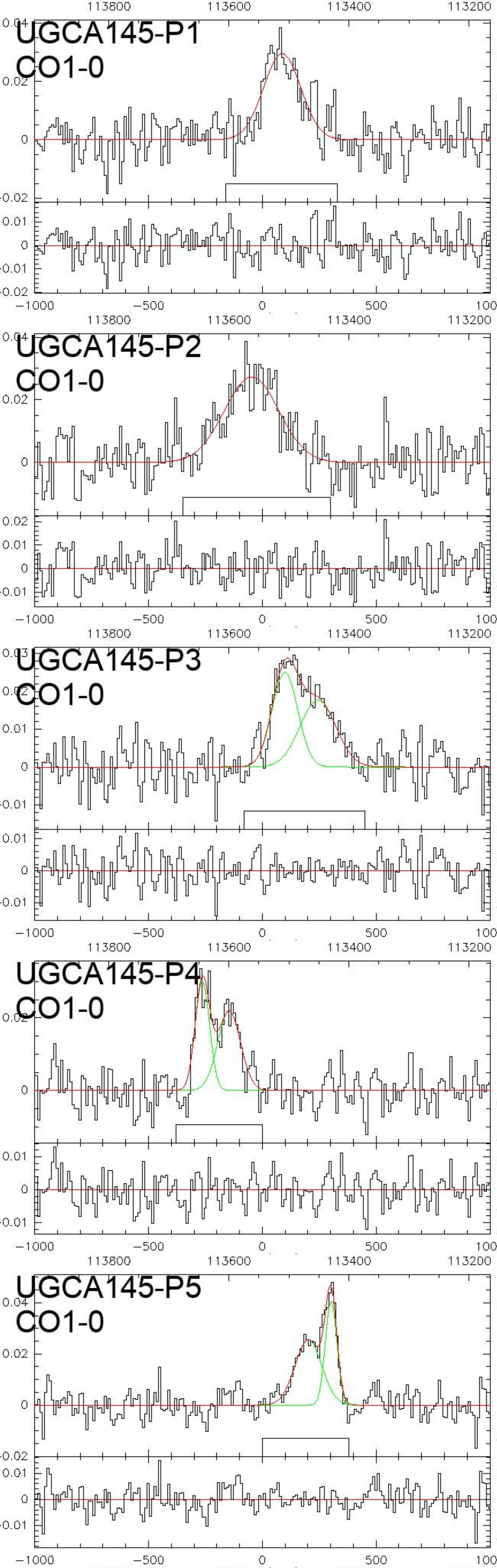}  
    \hfill
    \includegraphics[viewport=0 0 650 2050,clip,width=0.22\textwidth]
    {all_145.pdf}  
    \hfill
    \includegraphics[viewport=650 2050 1300 4100,clip,width=0.22\textwidth]
    {all_145.pdf} 
    \hfill
    \includegraphics[viewport=650 0 1300 2050,clip,width=0.22\textwidth]
    {all_145.pdf} 
\\

    \textbf{ $\mathrm{Velocity~(km/s)}$}
    \caption{Spectra of two CO molecular lines in UGCA 145. The $y$-axis in the upper half of each panel is the main beam temperature without main beam and forward efficiency corrections. The lower half is the residual of the best-fit. For all galaxies, the bottom $x$-axis displays the velocity scale ranging from $-1000$ to $+1000~\rm km~s^{-1}$. The top $x$-axis marks the corresponding frequency ranges of the emission lines, centered at the rest-frame frequencies determined by each galaxy's systemic velocity. All spectra are binned to a velocity resolution of $10~\rm km~s^{-1}$. The red line shows the overall fit, with green curves denoting the two Gaussian components where applicable. The black box outlines the velocity range determined by the line widths of $^{12}\mathrm{CO}~J=1-0$ emission.}\label{fig:spectra145}
\end{figure*}
%The lower $x$-axis for all galaxies shows the same velocity range within $\pm1000\rm~km~s^{-1}$. The top $x$-axis indicates the frequency range for the three different molecular lines of each galaxy, set according to the zero velocity, which is defined by the galaxy's systematic velocity \citep{Irwin12a}. All spectra are binned to a velocity resolution of $10\rm~km~s^{-1}$.

\newpage

\begin{figure*}[!ht]
    \centering
    \textbf{ $\mathrm{Frequency~(Hz)}$} \\ 
    \raisebox{3cm}{\rotatebox{90}{\centering \textbf{T (K)}}}\hspace{1em}
    \includegraphics[viewport=0 1230 650 2460,clip, width=0.22\textwidth]{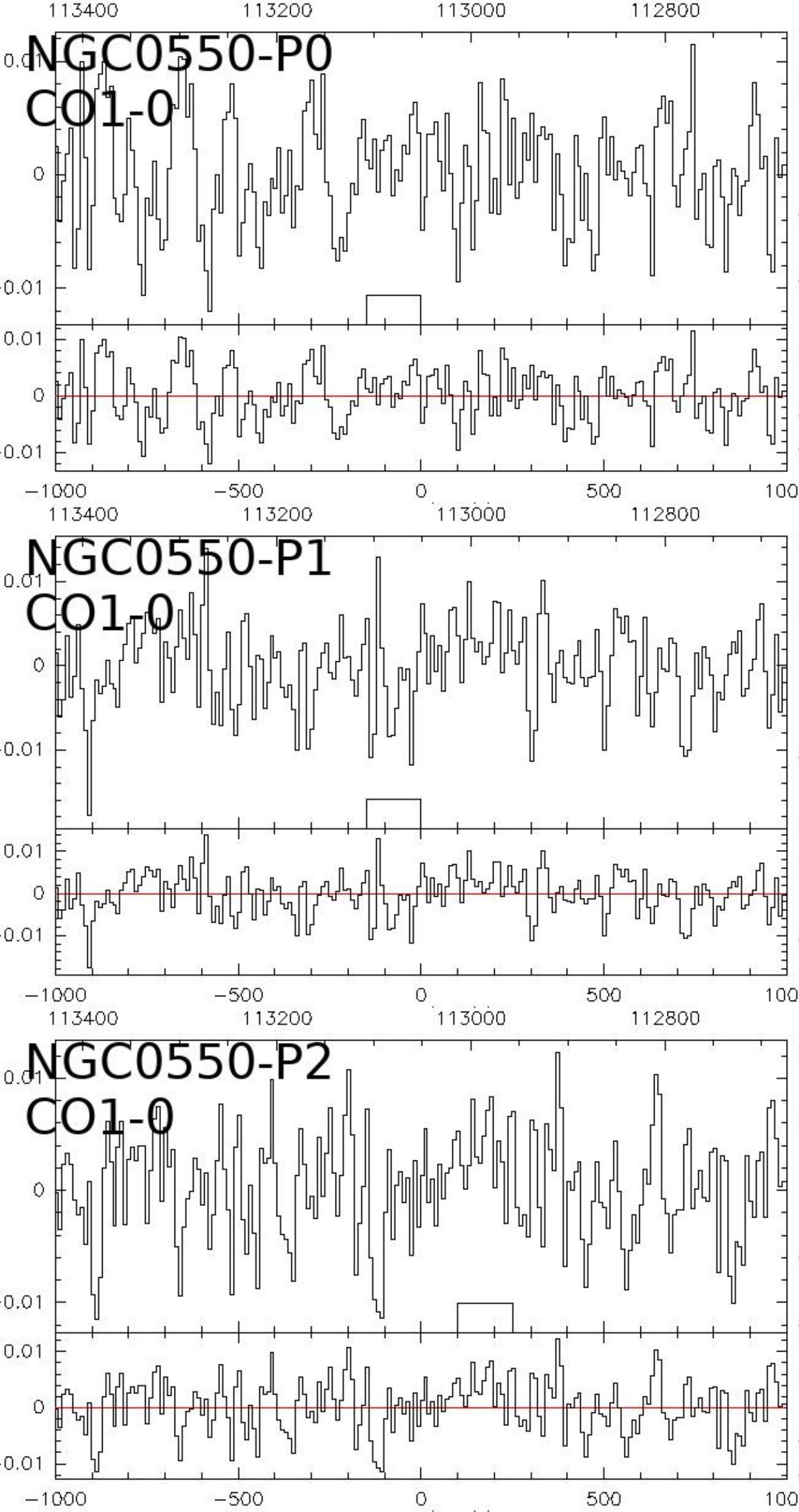}
    \hfill
    \includegraphics[viewport=0 0 650 1230,clip, width=0.22\textwidth]{all_550.pdf}  
    \hfill
    \includegraphics[viewport=650 1230 1300 2460,clip, width=0.22\textwidth]{all_550.pdf}  
    \hfill
    \includegraphics[viewport=650 0 1300 1230,clip, width=0.22\textwidth]{all_550.pdf}  
\\
    \textbf{ $\mathrm{Velocity~(km/s)}$}
    \caption{Spectra of two CO molecular lines in NGC 550. Similar as Figure~\ref{fig:spectra145}}\label{fig:spectra550}
\end{figure*}

\clearpage

\begin{figure*}[!ht]
    \centering

    \textbf{ $\mathrm{Frequency~(Hz)}$} \\
    \raisebox{6.2cm}{\rotatebox{90}{\centering \textbf{T (K)}}}\hspace{1em}
    \includegraphics[viewport=0 1640 650 3690,clip,width=0.22\textwidth]{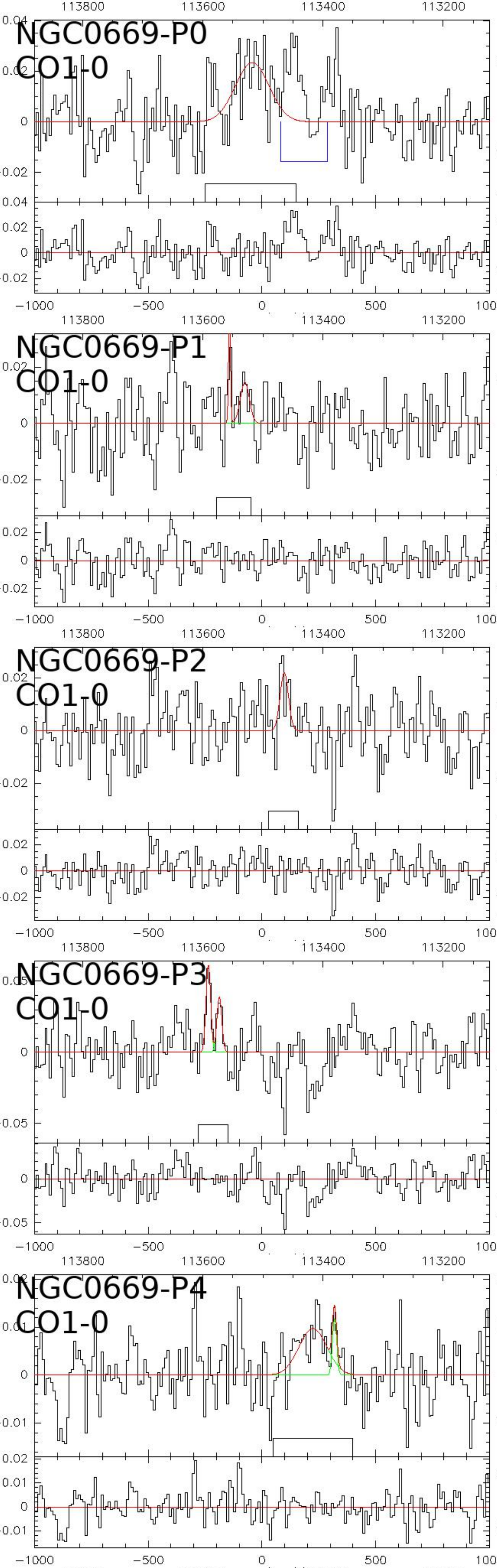} 
    \hfill
    \raisebox{2.5cm}{\includegraphics[viewport=0 0 650 1640,clip,width=0.22\textwidth]{all_669.pdf}}
    \hfill
    \includegraphics[viewport=650 1640 1300 3690,clip,width=0.22\textwidth]{all_669.pdf}
    \hfill
    \raisebox{2.5cm}{\includegraphics[viewport=650 0 1300 1640,clip,width=0.22\textwidth]{all_669.pdf}}
\\
    \textbf{ $\mathrm{Velocity~(km/s)}$}
    \caption{Spectra of two CO molecular lines in NGC 669. Similar as Figure~\ref{fig:spectra145}}\label{fig:spectra669}
\end{figure*}

\clearpage

\begin{figure*}[!ht]
    \centering

    \textbf{ $\mathrm{Frequency~(Hz)}$}\\
    \raisebox{5cm}{\rotatebox{90}{\centering \textbf{T (K)}}}\hspace{1em}
    \includegraphics[viewport=0 1230 650 2870,clip,width=0.22\textwidth]{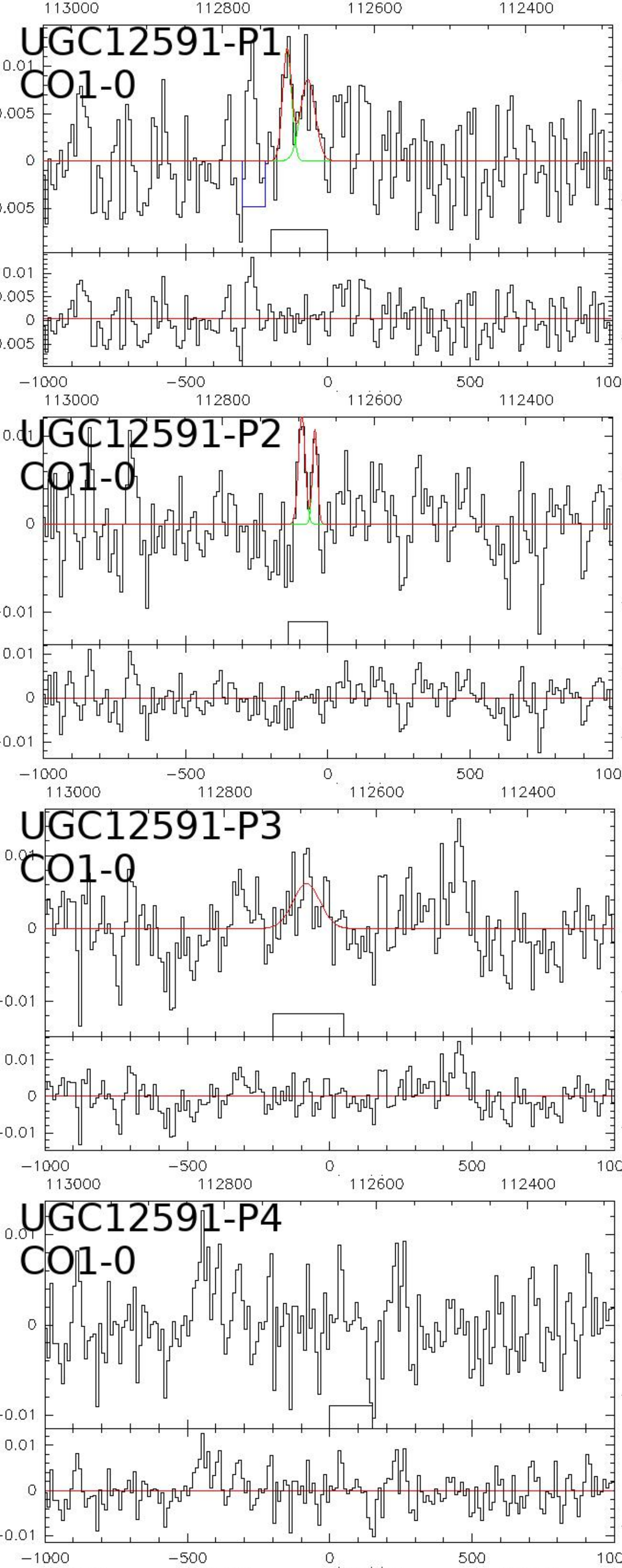} 
    \hfill
    \raisebox{2.5cm}{\includegraphics[viewport=0 0 650 1230,clip,width=0.22\textwidth]{all_12591.pdf}}
    \hfill
    \includegraphics[viewport=650 1230 1300 2870,clip,width=0.22\textwidth]{all_12591.pdf}
    \hfill
    \raisebox{2.5cm}{\includegraphics[viewport=650 0 1300 1230,clip,width=0.22\textwidth]{all_12591.pdf}}
\\
    \textbf{ $\mathrm{Velocity~(km/s)}$}
    \caption{Spectra of two CO molecular lines in UGC 12591. Similar as Figure~\ref{fig:spectra145}}\label{fig:spectra12591}
\end{figure*}

\bibliography{references}{}     

\end{document}